\documentclass[12pt,american,english, journal,draftclsnofoot,onecolumn]{IEEEtran}
\usepackage[T1]{fontenc}
\usepackage[latin9]{inputenc}
\usepackage{color}
\usepackage{float}
\usepackage{amsthm}
\usepackage{amsmath}
\usepackage{amssymb}
\usepackage{stackrel}
\usepackage{graphicx}
\usepackage{setspace}
\usepackage{esint}
\makeatletter

\floatstyle{ruled}
\newfloat{algorithm}{tbp}{loa}
\providecommand{\algorithmname}{Algorithm}
\floatname{algorithm}{\protect\algorithmname}

\theoremstyle{plain}
\newtheorem{thm}{\protect\theoremname}
\theoremstyle{plain}
\newtheorem{lem}[thm]{\protect\lemmaname}

\usepackage{colortbl}
\usepackage{cite}
\usepackage{nomencl}
\usepackage{flushend}
\usepackage{stfloats}
\usepackage[margin=8pt,font=footnotesize,labelfont=md,labelsep=colon]{caption}

\@ifundefined{showcaptionsetup}{}{%
 \PassOptionsToPackage{caption=false}{subfig}}
\usepackage{subfig}
\makeatother

\usepackage{babel}
\addto\captionsamerican{\renewcommand{\algorithmname}{Algorithm}}
\addto\captionsamerican{\renewcommand{\lemmaname}{Lemma}}
\addto\captionsamerican{\renewcommand{\theoremname}{Theorem}}
\addto\captionsenglish{\renewcommand{\lemmaname}{Lemma}}
\addto\captionsenglish{\renewcommand{\theoremname}{Theorem}}
\providecommand{\lemmaname}{Lemma}
\providecommand{\theoremname}{Theorem}

\begin{document}

\title{Rate Splitting with Finite Constellations: The Benefits of Interference
Exploitation vs Suppression}

\author{Abdelhamid Salem,\textit{\normalsize{} Member, IEEE}{\normalsize{},}
Christos Masouros, \textit{\normalsize{}Senior Member, IEEE}, and
Bruno Clerckx, \textit{\normalsize{}Senior Member, IEEE}\\
\thanks{Abdelhamid Salem and Christos Masouros are with the department of
Electronic and Electrical Engineering, University College London,
London, UK, (emails: \{a.salem, c.masouros\}@ucl.ac.uk). 

B. Clerckx is with the Electrical and Electronic Engineering Department,
Imperial College London, London SW7 2AZ, U.K. (e-mail: b.clerckx@imperial.ac.uk).

Parts of this paper were presented at the WCNC 2019 \cite{WCNC19}.

This work was supported in part by the U.K. Engineering and Physical
Sciences Research Council (EPSRC) under grants EP/R007934/1, in part
by the U.K. Engineering and Physical Sciences Research Council (EPSRC)
under grants EP/N015312/1 and EP/R511547/1.%
} }

\maketitle
\selectlanguage{american}%
\thispagestyle{empty}
\selectlanguage{english}%
\begin{abstract}
Rate-Splitting (RS) has been proposed recently to enhance the performance
of multi-user multiple-input multiple-output (MU-MIMO) systems. In
RS, a user message is split into a common and a private part, where
the common part is decoded by all users, while the private part is
decoded only by the intended user. In this paper, we study RS under
a phase-shift keying (PSK) input alphabet for multi-user multi-antenna
system and propose a constructive interference (CI) exploitation approach
to further enhance the sum-rate achieved by RS under PSK signaling.
To that end, new analytical expressions for the ergodic sum-rate are
derived for two precoding techniques of the private messages, namely,
1) a traditional interference suppression zero-forcing (ZF) precoding
approach, 2) a closed-form CI precoding approach. Our analysis is
presented for perfect channel state information at the transmitter
(CSIT), and is extended to imperfect CSIT knowledge. A novel power
allocation strategy, specifically suited for the finite alphabet setup,
is derived and shown to lead to superior performance for RS over conventional
linear precoding not relying on RS (NoRS). The results in this work
validate the significant sum-rate gain of RS with CI over the conventional
RS with ZF and NoRS. \end{abstract}


\section{Introduction}

\selectlanguage{american}%
\IEEEPARstart{T}{he}\foreignlanguage{english}{ recent years have
witnessed the widespread application of multi-user multiple-input
multiple-output (MU-MIMO) systems, due to their reliability and high
spectral efficiency \cite{MIMO1,MIMO2,MIMO3}. However, in practical
communication networks, the advantages of MU-MIMO systems are often
impacted by interference \cite{MIMO1,MIMO2,MIMO3}. Consequently,
a considerable amount of researches has focused on improving the performance
of MU-MIMO systems \cite{MIMO3,pdfZF,Twway}. In this regard, Rate-Splitting
(RS) approach was recently proposed and investigated in different
scenarios to enhance the performance of MU-MIMO systems \cite{RS1,Rs3,Rs2,Rs4,RS5}.
RS scheme splits the users' messages into a common message and private
messages, and superimposes the common message on top of the private
messages. Using Successive Interference Cancellation (SIC) at the
receivers, the common message is first decoded by all the users, and
each private message is then decoded only by its intended user. By
adjusting the message split and the power allocated to the common
and private messages, RS has the ability to better handle the multiuser
interference. RS has been studied in multiuser multi-antenna setups
with both perfect and imperfect CSIT. In \cite{Rs2}, authors analyzed
the sum-rate gain achieved by RS over conventional multi-user linear
precoding (NoRS) in a two-user multi-antenna broadcast channel with
imperfect CSIT, and considered that the common message is transmitted
via a space and space-time design. In \cite{Rs2,Rs3,Rs4,RS5}, again
considering imperfect CSIT, the authors leveraged convex optimization
to optimize the precoders of the common and private messages to maximize
the ergodic sum-rate and the max-min rate, respectively, and again
showed the superiority of RS over NoRS. In \cite{Rs6}, RS was designed
and its performance analyzed for Massive MIMO with imperfect CSIT
and shown to outperform the conventional NoRS approach. In \cite{RS5},
a multi-pair Massive MIMO relay system with imperfect CSIT was considered
and RS was shown to lead to higher robustness compared to NoRS. In
\cite{newrs2}, RS was designed for a multi-antenna multi-cell system
with imperfect CSIT, and showed the superiority in a Degrees-of-Freedom
sense over NoRS. The benefits of RS have also been highlighted in
multiuser multi-antenna system with perfect CSIT as in \cite{BrunoA,BrunoB},
and performance gains were highlighted over both NoRS and power-domain
Non-Orthogonal Multiple Access (NOMA) techniques.}

\selectlanguage{english}%
Another line of research has recently proposed constructive interference
(CI) precoding techniques to enhance the performance of downlink MU-MIMO
systems \cite{CI1,A,CI2,CI3}. In contrast to the conventional interference
mitigation techniques, where the knowledge of the interference is
used to cancel it, the main idea of the CI is to use the interference
to improve the system performance. The CI precoding technique exploits
interference that can be known to the transmitter to increase the
useful signal received power \cite{CI1,A,CI2,CI3}. That is, with
the knowledge of the CSI and users\textquoteright{} data symbols,
the interference can be classified as constructive and destructive.
The interference signal is considered to be constructive to the transmitted
signal if it pushes/moves the received symbols away from the decision
thresholds of the constellation towards the direction of the desired
symbol. Accordingly, the transmit precoding can be designed such that
the resulting interference is constructive to the desired symbol. 

The concept of CI has been extensively studied in literature. This
line of work has been introduced in \cite{CI1}, where the CI precoding
scheme for the downlink of PSK-based MIMO systems has been proposed.
In this work, it was shown that the effective signal to interference-plus-noise
ratio (SINR) can be enhanced without the need to increase the transmitted
signal power at the base station (BS). In \cite{A}, an optimization-based
precoder in the form of pre-scaling has been designed for the first
time using the concept of CI. Thereof, \cite{CI2} proposed transmit
beamforming schemes for the MU-MIMO downlink that minimize the transmit
power for generic PSK signals. In \cite{Luxm0}, a transmission algorithm
that exploits the constructive multiuser interference was proposed.
The authors in \cite{Luxm1,Luxm2} studied a general category of CI
regions, namely distance preserving CI region, where the full characterization
for a generic constellation was provided. In \cite{CI3,CI5}, CI precoding
scheme was applied in wireless power transfer scenario in order to
minimize the transmit power while guaranteeing the energy harvesting
and the quality of service (QoS) constraints for PSK messages. Further
work in \cite{CI4} applied the CI concept to Massive MIMO systems.
Very recently, the authors in \cite{angLi} derived closed-form precoding
expression for CI exploitation in the MU-MIMO downlink. The closed-form
precoder in this work has for the first time made the application
of CI exploitation practical, and has further paved the way for the
development of communication theoretic analysis of the benefits of
CI. While the above literature has addressed traditional downlink
transmission, the application of the CI concept to RS approaches remains
an open problem, due to the finite constellation input that CI requires.

Accordingly, in this paper, we provide the first attempt to combine
those two lines of research on RS and CI, and employ the CI precoding
technique to further enhance the sum-rate achieved by RS scheme in
MU-MIMO systems under a PSK input alphabet%
\footnote{We note that, while traditional analysis focus on Gaussian signaling,
the study of finite constellation signaling is of particular importance,
since finite constellations are applied in practice.%
}. In this regard and in order to provide fair comparison, new analytical
expressions for the ergodic sum-rate are derived for two precoding
techniques of the private messages, namely, 1) a closed-form CI precoding
approach, 2) a traditional interference suppression zero-forcing (ZF)
precoding approach. Our analysis is presented for perfect channel
state information at the BS (CSIT), and extended to imperfect CSIT.
Additionally, the conventional transmission, NoRS, is also studied
in this paper. Furthermore, a power allocation scheme that can achieve
superiority of RS over the NoRS in finite alphabet systems is proposed
and investigated.

For clarity we list the major contributions of this work as follows.
\begin{enumerate}
\item First, new analytical expressions for the ergodic sum-rate are derived
for RS based on finite constellations with CI and ZF precoding schemes
for the private messages. Both perfect CSIT and imperfect CSIT are
considered. This contrasts with the existing literature that either
study NoRS based on finite constellation with CI/ZF precoding, or
RS based on Gaussian inputs. This is the first paper that a) studies
RS with finite constellations, b) combines RS with CI precoding.
\item Second, a novel power allocation algorithm is introduced to optimize
the resulting sum-rate in the finite alphabet scenario.
\item Third, Monte-Carlo simulations are provided to confirm the analysis,
and the impact of the different system parameters on the achievable
sum-rate are examined and investigated. 
\end{enumerate}
The results in this work show clearly that the sum-rate of RS with
CI outperforms the sum rate of RS with ZF and NoRS (with either ZF
or CI) transmission techniques.

Notations: $h$, $\mathbf{h}$, and \textbf{$\mathbf{H}$} denote
a scalar, a vector and a matrix, respectively. $(\cdot)^{H}$, $(\cdot)^{T}$
and $\textrm{diag}\left(.\right)$ denote conjugate transposition,
transposition and diagonal of a matrix, respectively. $\mathcal{E}\left[.\right]$
denotes average operation. $\left[\mathbf{h}\right]_{k}$ denotes
the $k^{th}$ element in $\mathbf{h}$, $\left|.\right|$ denotes
the absolute value, , and $\left\Vert .\right\Vert ^{2}$ denotes
the second norm. $\mathbb{C}^{K\times N}$ represents an $K\times N$
matrix, and $\mathbf{I}$ denotes the identity matrix.

\section{System Model}

We consider a MU-MIMO system, in which an $N-$antennas BS node communicates
with $K$-single antenna users in a downlink scenario using the RS
strategy. In this system the channels are assumed to be independent
identically distributed (i.i.d) Rayleigh fading channels. The channel
matrix between the BS and the $K$ users is denoted by $\mathbf{H}\in\mathbb{C}^{K\times N}$,
which can be written as $\mathbf{H}=\mathbf{D}^{1/2}\boldsymbol{\textrm{G}}$
where $\boldsymbol{\textrm{G}}\in\mathbb{C}^{K\times N}$ contains
i.i.d $\mathcal{CN}\left(\text{0,}1\right)$ elements represent small-scale
fading coefficients and $\mathbf{D}\in\mathbb{C}^{K\times K}$ is
a diagonal matrix represents the path-loss attenuation with $\left[\mathbf{D}\right]_{kk}=d_{k}^{-m}$
,  where $d_{k}$ is the distance between the BS and the $k^{th}$
user, and $m$ is the path loss exponent. It is also assumed that
the signal is equiprobably drawn from an $M$-PSK constellation. 

Therefore, the BS transmits $K$ independent messages $Q_{t,1},...,Q_{t,K}$
uniformly drawn from the sets $\mathcal{Q}_{t,1},...,\mathcal{Q}_{t,K}$,
and intended for users $1,...,K$ respectively. In RS, each user message
is split into a common part and a private part, i.e., $Q_{t,1}=\left\{ Q_{c,k},Q_{p,k}\right\} $%
\footnote{The subscript $t$ here denotes total, which is explained that $Q_{t,k}$
is composed of two parts. The subscripts $c$ and $p$ are used for
common part and private parts, respectively.%
} with $Q_{c,k}\in\mathcal{Q}_{c,k}$, $Q_{p,k}\in\mathcal{Q}_{p,k}$,
and $\mathcal{Q}_{c,k}\times\mathcal{Q}_{p,k}=\mathcal{Q}_{t,k}$.
The common message is composed by packing the common parts such that
$Q_{c}=\left\{ Q_{c,1},...,Q_{c,K}\right\} \in\mathcal{Q}_{c,1}\times...\times\mathcal{Q}_{c,K}$.
The resulting $K+1$ messages are encoded into the independent data
streams $x_{c},x_{1},....,x_{K}$, where $x_{c}$ and $x_{k}$ represent
the encoded common and private symbols \cite{Rs3}. The $K+1$ symbols
are grouped in a signal vector $\mathbf{x}=\left[x_{c},x_{1},....,x_{K}\right]^{T}\in\mathbb{C}^{K+1}$,
where $\mathcal{E}\left\{ \mathbf{x}\mathbf{x}^{H}\right\} =\textrm{I}$.
Then the symbols are mapped to the BS antennas by a linear precoding
matrix defined as $\mathbf{W}=\left[\mathbf{w}_{c},\mathbf{w}_{1},....\mathbf{w}_{K}\right]$
where $\mathbf{w}_{c}\in\mathbb{C}^{N}$ denotes the common precoder
and $\mathbf{w}_{k}\in\mathbb{C}^{N}$ is the $k^{th}$ private precoder.
Therefore, the transmitted signal can be mathematically expressed
by \cite{RS1,Rs2,Rs3}

\begin{equation}
\mathbf{s}=\mathbf{W}\mathbf{x}=\sqrt{P_{c}}\mathbf{w}_{c}x_{c}+\stackrel[k=1]{K}{\sum}\sqrt{P_{p}}\mathbf{w}_{k}x_{k}
\end{equation}

\noindent where $\mathbf{W}=\left[\mathbf{w}_{c},\mathbf{w}_{1},....\mathbf{w}_{K}\right]$,$\mathbf{w}_{c}$
denotes the common precoder of the common message and $\mathbf{w}_{k}$
is the $k^{th}$ private precoder. In addition, $P_{c}$ and $P_{p}$
are the power allocated to the common message and the power allocated
to the private message, respectively, where $P_{c}=\left(1-t\right)P$
and $P_{p}=\dfrac{tP}{K}$, $0\leq t\leq1$ and $P$ is the total
power%
\footnote{We assume a uniform power allocation among all the private symbols,
similarly to other works on RS \cite{Rs2,Rs6}. Although this assumption
does not produce the optimal performance, it allows us to find tractable
results. This assumption is commonly used in practice, e.g. LTE and
LTE-A. %
}. Conventional multi-user linear precoding without RS, NoRS, is a
particular instance of the RS strategy and is obtained by turning
of the common message and allocating all transmit power exclusively
to the privates messages. The received signal at the $k^{th}$ user
in this system can be written as

\begin{equation}
y_{k}=\mathbf{h}_{k}\mathbf{W}\mathbf{x}+n_{k}=\sqrt{P_{c}}\mathbf{h}_{k}\mathbf{w}_{c}x_{c}+\stackrel[k=1]{K}{\sum}\sqrt{P_{p}}\mathbf{h}_{k}\mathbf{w}_{k}x_{k}+n_{k}
\end{equation}

\noindent where $\mathbf{h}_{k}$ is the channel vector from the BS
to user $k$, $n_{k}$ is the additive wight Gaussian noise (AWGN)
at the $k^{th}$ user, $n_{k}\sim\mathcal{CN}\left(\text{0, }\sigma_{k}^{2}\right)$.
At the user side, the common symbol is decoded firstly by treating
the interference from the private messages as noise, and then each
user decodes its own message after canceling the common message using
SIC technique. Therefore, after perfectly removing the contribution
from the common message, the received signal at the $k^{th}$ user
in this system can be written as

\begin{equation}
y_{k}^{p}=\mathbf{h}_{k}\mathbf{W}^{p}\mathbf{x}^{p}+n_{k}=\stackrel[k=1]{K}{\sum}\sqrt{P_{p}}\mathbf{h}_{k}\mathbf{w}_{k}x_{k}+n_{k}
\end{equation}

\noindent where $\mathbf{x}^{p}=\left[x_{1},....,x_{K}\right]^{T}$
and $\mathbf{W}^{p}=\left[\mathbf{w}_{1},....\mathbf{w}_{K}\right]$.
The sum rate in this scenario can be expressed by

\begin{equation}
R=R^{c}+\stackrel[k=1]{K}{\sum}R_{k}^{p}\label{eq:5}
\end{equation}

\noindent where $R^{c}$ is the rate for the common part, $R^{c}=\min\left(R_{1}^{c},R_{2}^{c},..,R_{k}^{c},..,R_{K}^{c}\right)$
, $R_{k}^{c}$ is the rate for the common message at user $k$, and
$R_{k}^{p}$ is the rate for the private part at the $k^{th}$ user. 

In this work, both perfect CSIT and imperfect CSIT are considered,
and delay-tolerant transmission is assumed. Hence the channel coding
can be achieved over a long sequence of channel states. Therefore,
transmitting the common and the $k^{th}$ private messages at ergodic
rates $\mathcal{E}\left\{ R_{k}^{c}\right\} $ and $\mathcal{E}\left\{ R_{k}^{p}\right\} $,
respectively, guarantees successful decoding by the $k^{th}$ user
\cite{RS5}. Hence, to guarantee the common message, $x_{c}$, is
successfully decoded and then canceled by the users, it should be
sent at an ergodic rate not exceeding $\min_{j}\left(\mathcal{E}\left\{ R_{j}^{c}\right\} \right)_{j=1}^{K}$
. Finally, the ergodic sum rate can be expressed by,

\begin{equation}
\mathcal{E}\left\{ R\right\} =\min_{j}\left(\mathcal{E}\left\{ R_{j}^{c}\right\} \right)_{j=1}^{K}+\stackrel[k=1]{K}{\sum}\mathcal{E}\left\{ R_{k}^{p}\right\} \label{eq:5-1}
\end{equation}

\section{Ergodic Sum Rate Analysis under PSK signaling and perfect CSIT}

In this scenario, the BS has perfect knowledge of the CSI, and the
precoding matrices have been designed based on this perfect knowledge.
Therefore, in this section two precoding techniques are considered.
In the first one, we use maximum ratio transmission (MRT) for the
common message and CI for the private messages, and in the second
one we use MRT for the common message and ZF for the private messages.

\subsection{RS: MRT/CI\label{sub:RS:-MRT/CI}}

In this scenario MRT technique is used for common message and CI precoding
for the private messages. Therefore, the precoder for the common and
the private messages can be written, respectively, as \cite{Rs6,angLi}

\begin{equation}
\mathbf{w}_{c}=\overset{K}{\underset{i=1}{\sum}}\beta_{c}\mathbf{h}_{i}^{H}
\end{equation}

\begin{equation}
\mathbf{W}_{CI}^{p}=\frac{1}{K}\beta_{p}\,\mathbf{H}^{H}\left(\mathbf{H}\mathbf{H}^{H}\right)^{-1}\textrm{diag}\left\{ \mathbf{V}^{-1}\mathbf{u}\right\} ,\label{eq:7}
\end{equation}

\noindent where $\beta_{c}=\frac{1}{\sqrt{\left\Vert \overset{K}{\underset{i=1}{\sum}}\mathbf{h}_{i}^{H}\right\Vert ^{2}}}$
and $\beta_{p}=\frac{1}{\sqrt{\mathbf{u}^{H}\mathbf{V}^{-1}\mathbf{u}}}$
are the scaling factor to meet the transmit power constraint at the
transmitter, while $\mathbf{V}=\textrm{diag}\left(\mathbf{x}^{pH}\right)\left(\mathbf{H}\mathbf{H}^{H}\right)^{-1}\textrm{diag}\left(\mathbf{x}^{p}\right)$
and $\mathbf{1}^{T}\mathbf{u}=1$. For simplicity and mathematical
tractability but without loss of generality, the normalization constants
$\beta_{c}$ and $\beta_{p}$ are designed to ensure that the long-term
total transmit power at the source is constrained, so it can be written
as \cite{Twway,angLi} $\beta_{c}=\frac{1}{\sqrt{\mathcal{E}\left\{ \left\Vert \overset{K}{\underset{i=1}{\sum}}\mathbf{h}_{i}^{H}\right\Vert ^{2}\right\} }}$
and $\beta_{p}=\frac{1}{\sqrt{\mathcal{E}\left\{ \mathbf{u}^{H}\mathbf{V}^{-1}\mathbf{u}\right\} }}$.
Since $\left\Vert \overset{K}{\underset{i=1}{\sum}}\mathbf{h}_{i}^{H}\right\Vert ^{2}$
and $\left(\mathbf{H}\mathbf{H}^{H}\right)$ has Gamma and Wishart
distributions respectively, we can find that, $\beta_{c}=\frac{1}{\sqrt{N\stackrel[i=1]{K}{\sum}}\varpi_{k}}$and
$\beta_{p}=\frac{1}{\sqrt{\mathbf{u}^{H}\textrm{diag}\left(\mathbf{x}^{H}\right)^{-1}\, N\mathbf{\Sigma}\left(\textrm{diag}\left(\mathbf{x}\right)\right)^{-1}\mathbf{u}}}$,
where$\Sigma=\textrm{D}$ and $\varpi_{k}=d_{k}^{-m}$ \cite{bookaspects}. 

\noindent From (\ref{eq:5-1}), we now need to calculate the ergodic
rate for the common and private messages as follows.

\subsubsection{\noindent Ergodic Rate for the Common Part}

\noindent The ergodic rate for the common part at user $k$ under
PSK signaling can be written as\cite{interference1,precoding6,precoding2},

\[
\mathcal{E}\left\{ R_{k}^{c}\right\} =\log_{2}M-\frac{1}{M^{N}}\stackrel[m=1]{M^{N}}{\sum}\underset{\varphi}{\underbrace{\mathcal{E}_{\mathbf{h},n_{k}}\left\{ \log_{2}\stackrel[i=1]{M^{N}}{\sum}e^{\frac{-\left|\mathbf{h}_{k}\mathbf{W}\mathbf{x}_{m,i}+n_{k}\right|^{2}}{\sigma_{k}^{2}}}\right\} }}
\]

\begin{equation}
+\frac{1}{M^{N-1}}\stackrel[m=1]{M^{N-1}}{\sum}\underset{\psi}{\underbrace{\mathcal{E}_{\mathbf{h},n_{k}}\left\{ \log_{2}\stackrel[i=1]{M^{N-1}}{\sum}e^{\frac{-\left|\mathbf{h}_{k}\mathbf{W}_{CI}^{p}\mathbf{x}_{m,i}^{p}+n_{k}\right|^{2}}{\sigma_{k}^{2}}}\right\} }},\label{eq:7-1}
\end{equation}

\noindent where $\mathbf{W}=\left[\mathbf{w}_{c},\mathbf{W}_{CI}^{p}\right]$,
$\mathbf{x}_{m,i}=\boldsymbol{\textrm{x}}_{m}-\mathbf{x}_{i}$, $\mathbf{x}_{m}$
and $\mathbf{x}_{i}$ contain $N$ symbols, which are taken from the
equiprobable constellation set with cardinality $M$%
\footnote{Each input $\mathbf{x}_{i}$ consists of symbols taken from the $M$-PSK
constellation.%
}.
\begin{IEEEproof}
The proof of the above follows known derivations from the finite constellation
rate analysis literature, and due to the paper length limitation,
the proof of (\ref{eq:7-1}) has been omitted in this paper. 
\end{IEEEproof}
Similar to the Gaussian input assumption case, (\ref{eq:7-1}) reveals
that the achievable rate suffering from the interference caused by
other signals. The first term in (\ref{eq:7-1}), $\varphi$, contains
all the received signals at user $k$, while the second term, $\psi$,
contains only the interference signals.

\noindent By invoking Jensen inequality, the first term in (\ref{eq:7-1}),
$\varphi$, can be expressed by

\begin{equation}
\varphi=\mathcal{E}_{\mathbf{h},n_{k}}\left\{ \log_{2}\stackrel[i=1]{M^{N}}{\sum}e^{\frac{-\left|\mathbf{h}_{k}\mathbf{W}\mathbf{x}_{m,i}+n_{k}\right|^{2}}{\sigma_{k}^{2}}}\right\} \leq\log_{2}\stackrel[i=1]{M^{N}}{\sum}\mathcal{E}_{\mathbf{h},n_{k}}\left\{ e^{\frac{-\left|\mathbf{h}_{k}\mathbf{W}\mathbf{x}_{m,i}+n_{k}\right|^{2}}{\sigma_{k}^{2}}}\right\} 
\end{equation}

Since the noise, $n_{k}$, has Gaussian distribution, the average
over the noise can be derived as

\begin{equation}
\mathcal{E}_{n_{k}}\left\{ e^{\frac{-\left|\mathbf{h}_{k}\mathbf{W}\mathbf{x}_{m,i}+n_{k}\right|^{2}}{\sigma_{k}^{2}}}\right\} =\frac{1}{\pi\sigma^{2}}\underset{n_{k}}{\int}e^{-\frac{\left|\mathbf{h}_{k}\mathbf{W}\mathbf{x}_{m,i}+n_{k}\right|^{2}+\left|n_{k}\right|^{2}}{\sigma_{k}^{2}}}dn_{k}.\label{eq:10}
\end{equation}

Using the integrals of exponential function in \cite{book2}, we can
find

\begin{equation}
\mathcal{E}_{n}\left\{ e^{\frac{-\left|\mathbf{h}_{k}\mathbf{W}\mathbf{x}_{m,i}+n_{k}\right|^{2}}{\sigma_{k}^{2}}}\right\} =e^{-\frac{\left|\mathbf{h}_{k}\mathbf{W}\mathbf{x}_{m,i}\right|^{2}}{2\sigma_{k}^{2}}}.\label{eq:12-1}
\end{equation}

Now, the average over the channel can be derived as

\begin{equation}
\mathcal{E}_{\mathbf{h}}\left\{ e^{-\frac{\left|\mathbf{h}_{k}\mathbf{W}\mathbf{x}_{m,i}\right|^{2}}{2\sigma_{k}^{2}}}\right\} =\mathcal{E}_{\mathbf{h}}\left\{ e^{-\frac{\left|\sqrt{P_{c}}\mathbf{h}_{k}\mathbf{w}_{c}\left[\mathbf{x}_{m,i}\right]_{1}+\sqrt{P_{p}}\mathbf{h}_{k}\mathbf{W}_{CI}^{p}\mathbf{x}_{m,i}^{p}\right|^{2}}{2\sigma_{k}^{2}}}\right\} 
\end{equation}

\noindent which can be written as 

\begin{equation}
\mathcal{E}_{\mathbf{h}}\left\{ e^{-\frac{\left|\mathbf{h}_{k}\mathbf{W}\mathbf{x}_{m,i}\right|^{2}}{2\sigma_{k}^{2}}}\right\} =\mathcal{E}_{\mathbf{\mathbf{h}}}\left\{ e^{-\frac{\left|\sqrt{P_{c}}\beta_{c}\left(\overset{K}{\underset{i=1}{\sum}}\mathbf{h}_{k}\mathbf{h}_{i}^{H}\right)\left[\mathbf{x}_{m,i}\right]_{1}+\left(\frac{\sqrt{P_{p}}\beta_{p}}{K}\mathbf{a}_{k}\left(\textrm{diag}\left(\mathbf{x}^{p}{}^{H}\right)\right)^{-1}\left(\mathbf{H}\mathbf{H}^{H}\right)\left(\textrm{diag}\left(\mathbf{x}^{p}\right)\right)^{-1}\mathbf{u}\left[\mathbf{x}_{m,i}^{p}\right]_{k}\right)\right|^{2}}{2\sigma_{k}^{2}}}\right\} ,\label{eq:12}
\end{equation}

\noindent where  $\mathbf{a}_{k}$ is a $1\times K$ vector all the
elements of this vector are zeros except the $k^{th}$ element is
one. Therefore, the first term $\varphi$ can be expressed as

\[
\varphi=\log_{2}\stackrel[i=1]{M^{N}}{\sum}
\]

\begin{equation}
\mathcal{E}_{\mathbf{h}}e^{-\frac{P\left|\xi\right|^{2}\left\Vert \mathbf{A}_{k}\right\Vert ^{2}\overset{\varPsi_{m,i}}{\overbrace{\left|\frac{\sqrt{\left(1-t\right)}\beta_{c}\left(\overset{K}{\underset{i=1}{\sum}}\mathbf{h}_{k}\mathbf{h}_{i}^{H}\right)\left[\mathbf{x}_{m,i}\right]_{1}}{\left|\xi\right|\left\Vert \mathbf{A}_{k}\right\Vert }+\frac{\sqrt{t}\beta_{p}\mathbf{a}_{k}\left(\textrm{diag}\left(\mathbf{x}^{p}{}^{H}\right)\right)^{-1}\mathbf{A}\left(\textrm{diag}\left(\mathbf{x}^{p}\right)\right)^{-1}\mathbf{u}\left[\mathbf{x}_{m,i}^{p}\right]_{k}}{\left|\xi\right|K\left\Vert \mathbf{A}_{k}\right\Vert }\right|^{2}}}}{2\sigma_{k}^{2}}}\label{eq:13}
\end{equation}

\noindent where $\mathbf{A}=\left(\mathbf{H}\mathbf{H}^{H}\right)$,
$\xi=\frac{\beta_{p}}{K}\mathbf{a}_{k}\left(\textrm{diag}\left(\mathbf{x}^{p}{}^{H}\right)\right)^{-1}\left(\Sigma\right)\left(\textrm{diag}\left(\mathbf{x}^{p}\right)\right)^{-1}\mathbf{u}$.
Now, we can simplify the last expression in (\ref{eq:13}) to

\begin{equation}
\varphi=\log_{2}\stackrel[i=1]{M^{N}}{\sum}\mathcal{E}_{\mathbf{h}}\left\{ e^{-\frac{P\left|\xi\right|^{2}\left\Vert \mathbf{A}_{k}\right\Vert ^{2}\varPsi_{m,i}}{2\sigma_{k}^{2}}}\right\} .
\end{equation}

\noindent The term $\varPsi_{m,i}$ in (\ref{eq:13}) can be reduced
to $\varPsi_{m,i}=N\left(\left|\left(\sqrt{1-t}\left[\mathbf{x}_{m,i}\right]_{1}\right)+\left(\sqrt{t}\varpi_{k}^{-1}\left[\mathbf{x}_{m,i}^{p}\right]_{k}\right)\right|^{2}\right).$
 Let $z_{k}=\left\Vert \mathbf{A}_{k}\right\Vert ^{2}$ which has
Gamma distribution, .i.e., $z_{k}\thicksim\Gamma\left(\upsilon_{k},\theta_{k}\right)$,
with $\upsilon_{k}=N\left(N+1\right)$ degrees of freedom, therefore
the average over \textbf{$\mathbf{h}$ }is the moment generating function
(MGF) of the term,\textbf{ $\frac{P\left|\xi\right|^{2}\left\Vert \mathbf{A}_{k}\right\Vert ^{2}\varPsi_{m,i}}{2\sigma_{k}^{2}}$
}which can be found easily as

\begin{equation}
\varphi=\log_{2}\stackrel[i=1]{M^{N}}{\sum}\left(1+\frac{P\left|\xi\right|^{2}\theta_{k}\varPsi_{m,i}}{2\sigma_{k}^{2}}\right)^{-\upsilon_{k}}.
\end{equation}

\noindent For the second term, $\psi$, similarly using Jensen inequality
we can write

\begin{equation}
\psi=\mathcal{E}_{\mathbf{h},n_{k}}\left\{ \log_{2}\stackrel[i=1]{M^{N-1}}{\sum}e^{\frac{-\left|\mathbf{h}_{k}\mathbf{W}_{CI}^{p}\mathbf{x}_{m,i}^{p}+n_{k}\right|^{2}}{\sigma_{k}^{2}}}\right\} \leq\log_{2}\stackrel[i=1]{M^{N-1}}{\sum}\mathcal{E}_{\mathbf{h},n_{k}}\left\{ e^{\frac{-\left|\mathbf{h}_{k}\mathbf{W}_{CI}^{p}\mathbf{x}_{m,i}^{p}+n_{k}\right|^{2}}{\sigma_{k}^{2}}}\right\} .
\end{equation}

\noindent Similarly as in (\ref{eq:10}), since $n_{k}$ has Gaussian
distribution, we can write $\psi$ as 

\begin{equation}
\psi=\log_{2}\stackrel[i=1]{M^{N-1}}{\sum}\mathcal{E}_{\mathbf{h}}\left\{ e^{-\frac{\left|\left(\frac{\beta_{p}}{K}\boldsymbol{a}\left(\mathbf{H}\mathbf{H}^{H}\right)\boldsymbol{b}\left[\mathbf{x}_{m,i}^{p}\right]_{k}\right)\right|^{2}}{2\sigma_{k}^{2}}}\right\} \label{eq:18}
\end{equation}

\noindent where $a=\mathbf{a}_{k}\left(\textrm{diag}\left(\mathbf{x}^{p}{}^{H}\right)\right)^{-1}$
and $b=\left(\textrm{diag}\left(\mathbf{x}^{p}\right)\right)^{-1}\mathbf{u}$.
It was shown that, $Y=\frac{\boldsymbol{a}\left(\mathbf{H}\mathbf{H}^{H}\right)\boldsymbol{b}}{\mathbf{a}\Sigma\boldsymbol{b}}$
has a Gamma distribution \cite{bookaspects}. Therefore we can rewrite
(\ref{eq:18}) as 

\begin{equation}
\psi=\log_{2}\stackrel[i=1]{M^{N-1}}{\sum}\mathcal{E}_{\mathbf{h}}\left\{ e^{-\frac{\left|c\, Y\left[\mathbf{x}_{m,i}^{p}\right]_{k}\right|^{2}}{2\sigma_{k}^{2}}}\right\} 
\end{equation}

\noindent where $c=\frac{\beta_{p}\mathbf{a}\Sigma\boldsymbol{b}}{K}$.
Therefore we can get,

\begin{equation}
\psi=\log_{2}\stackrel[i=1]{M^{N-1}}{\sum}\stackrel[0]{\infty}{\int}e^{-\frac{\left|c\, Y\left[\mathbf{x}_{m,i}^{p}\right]_{k}\right|^{2}}{2\sigma_{k}^{2}}}\frac{e^{-Ky}\left(Ky\right)^{N-K}K}{\left(N-K\right)!}dy,
\end{equation}

\noindent which can be obtained as 

\[
\psi=\log_{2}\stackrel[i=1]{M^{N-1}}{\sum}\left(\left(\frac{2^{\left(\frac{1}{2}\left(N-K-1\right)\right)}K^{\left(N-K+1\right)}\left|\left[\mathbf{x}_{m,i}^{p}\right]_{k}\right|^{-2+K-N}}{\left(N-K\right)!}\right)\left(\left(\frac{c^{2}}{\sigma_{k}^{2}}\right)^{\frac{1}{2}\left(K-N-1\right)}\right)\right)
\]

\[
\times\left(\left(c^{2}\left|\left[\mathbf{x}_{m,i}^{p}\right]_{k}\right|\right)\Gamma\left(\frac{1}{2}\left(N-K+1\right)\right)\textrm{\ensuremath{_{1}\textrm{F}_{1}}}\left(\frac{1}{2}\left(N-K+1\right),\frac{1}{2},\frac{K^{2}\sigma_{k}^{2}}{2c^{2}\left|\left[\mathbf{x}_{m,i}^{p}\right]_{k}\right|^{2}}\right)\right.
\]

\begin{equation}
\left.\left.-\sqrt{2}K\, c\,\sigma_{k}^{2}\Gamma\left(\frac{1}{2}\left(N-K+2\right)\right){}_{1}\textrm{F}_{1}\left(\frac{1}{2}\left(N-K+2\right),\frac{3}{2},\frac{K^{2}\sigma_{k}^{2}}{2c^{2}\left|\left[\mathbf{x}_{m,i}^{p}\right]_{k}\right|^{2}}\right)\right)\right).\label{eq:23}
\end{equation}

\noindent where $\textrm{\ensuremath{_{1}\textrm{F}_{1}}}$ is the
Hypergeometric function.

It is noted that Jensen\textquoteright s inequality has been used
in the two terms in (\ref{eq:7-1}). Accordingly, the resulting expression
cannot lead to a strict bound on the resulting rate. Nevertheless,
since the involved rate is based on a finite constellation, the resulting
low-SNR and high-SNR approximation match the exact rate. In the intermediate
SNR regions, it can be observed that the bounding errors of the two
terms have similar values which results in an accurate overall approximation,
as already verified in relevant analysis in \cite{precoding6}. We
note that the rate approximations show a very close match to our Monte
Carlo simulations in our results of Section VII.

\subsubsection{\noindent Ergodic Rate for the Private Part}

\noindent The ergodic rate for the private part at user $k$ under
PSK signaling, using CI precoding technique can be written as\cite{interference1,precoding6,precoding2},

\begin{equation}
\mathcal{E}\left\{ R_{k}^{p}\right\} =\log_{2}M-\,\frac{1}{M^{N-1}}\stackrel[m=1]{M^{N-1}}{\sum}\underset{\psi}{\underbrace{\mathcal{E}\left\{ \log_{2}\stackrel[i=1]{M^{N-1}}{\sum}e^{\frac{-\left|\mathbf{h}_{k}\mathbf{W}_{CI}^{p}\mathbf{x}_{m,i}^{p}\right|^{2}}{2\sigma_{k}^{2}}}\right\} }}.\label{eq:23-1}
\end{equation}

\noindent   By using Jensen inequality, and following similar steps
as in the previous sub-section we can find the average of the term
$\psi$ in (\ref{eq:23-1}) as in (\ref{eq:23}).

\subsection{RS: MRT/ZF\label{sub:RS:-MRT/ZF}}

In this case we implement MRT technique for common signal and ZF precoding
for the private messages. Therefore, the precoding for the common
and the private messages can be written, respectively, as \cite{Rs6,angLi}

\begin{equation}
\mathbf{w}_{c}=\overset{K}{\underset{i=1}{\sum}}\beta_{c}\mathbf{h}_{i}^{H}\label{eq:27}
\end{equation}

\begin{equation}
\mathbf{W}_{ZF}^{p}=\beta_{p}\,\mathbf{H}^{H}\left(\mathbf{H}\mathbf{H}^{H}\right)^{-1},\label{eq:24}
\end{equation}

\noindent where $\beta_{c}$ and $\beta_{p}$ are the scaling factors
to meet the transmit power constraint at the transmitter, which can
be expressed as $\beta_{c}=\frac{1}{\sqrt{\left\Vert \overset{K}{\underset{i=1}{\sum}}\mathbf{h}_{i}^{H}\right\Vert ^{2}}}$
and $\beta_{p}=\sqrt{\frac{1}{\mathbf{x}^{H}\left(\mathbf{H}\mathbf{H}^{H}\right)^{-1}\mathbf{x}}}$.
Similarly as in the MRT/CI scenario, and for mathematical tractability
but without loss of generality, the normalization constants $\beta_{c}$
and $\beta_{p}$ are designed to ensure that the long-term total transmit
power at the source is constrained, so it can be written as \cite{Twway,angLi},
$\beta_{c}=\frac{1}{\sqrt{\mathcal{E}\left\{ \left\Vert \overset{K}{\underset{i=1}{\sum}}\mathbf{h}_{i}^{H}\right\Vert ^{2}\right\} }}$
and $\beta_{p}=\frac{1}{\sqrt{\mathcal{E}\left\{ \mathbf{x}^{H}\left(\mathbf{H}\mathbf{H}^{H}\right)^{-1}\mathbf{x}\right\} }}$,
respectively. Since $\left\Vert \overset{K}{\underset{i=1}{\sum}}\mathbf{h}_{i}^{H}\right\Vert ^{2}$
and $\frac{1}{\mathbf{s}^{H}\left(\mathbf{H}\mathbf{H}^{H}\right)^{-1}\mathbf{s}}$
both have Gamma distribution \cite{pdfZF,CI1}, we can find that,
$\beta_{c}=\frac{1}{\sqrt{N\stackrel[i=1]{K}{\sum}}\varpi_{k}}$and
$\beta_{p}=\sqrt{\frac{\Gamma\left(2-K+N\right)}{\left(K\left(N-K\right)!\right)}}$
\cite{bookaspects}.

\subsubsection{\noindent Ergodic Rate for the Common Part}

The ergodic rate for the common part at user k can be written as

\[
\mathcal{E}\left\{ R_{k}^{c}\right\} =\log_{2}M-\frac{1}{M^{N}}\stackrel[m=1]{M^{N}}{\sum}\underset{\varphi}{\underbrace{\mathcal{E}_{\mathbf{h},n_{k}}\left\{ \log_{2}\stackrel[i=1]{M^{N}}{\sum}e^{\frac{-\left|\mathbf{h}_{k}\mathbf{W}\mathbf{x}_{m,i}+n_{k}\right|^{2}}{\sigma_{k}^{2}}}\right\} }}
\]

\begin{equation}
+\frac{1}{M^{N-1}}\stackrel[m=1]{M^{N-1}}{\sum}\underset{\psi}{\underbrace{\mathcal{E}_{\mathbf{h},n_{k}}\left\{ \log_{2}\stackrel[i=1]{M^{N-1}}{\sum}e^{\frac{-\left|\mathbf{h}_{k}\mathbf{W}_{ZF}^{p}\mathbf{x}_{m,i}^{p}+n_{k}\right|^{2}}{\sigma_{k}^{2}}}\right\} }},\label{eq:7-1-1}
\end{equation}

\noindent where $\mathbf{W}=\left[\mathbf{w}_{c},\mathbf{W}_{ZF}^{p}\right]$.
By using Jensen inequality, the first term in (\ref{eq:7-1-1}),
$\varphi$, can be written as

\begin{equation}
\varphi=\mathcal{E}_{\mathbf{h},n_{k}}\left\{ \log_{2}\stackrel[i=1]{M^{N}}{\sum}e^{\frac{-\left|\mathbf{h}_{k}\mathbf{W}\mathbf{x}_{m,i}+n_{k}\right|^{2}}{\sigma_{k}^{2}}}\right\} \leq\log_{2}\stackrel[i=1]{M^{N}}{\sum}\mathcal{E}_{\mathbf{h},n_{k}}\left\{ e^{\frac{-\left|\mathbf{h}_{k}\mathbf{W}\mathbf{x}_{m,i}+n_{k}\right|^{2}}{\sigma_{k}^{2}}}\right\} .
\end{equation}

Since the noise $n_{k}$ is Gaussian distributed, using the integrals
of exponential function in \cite{book2} the average over the noise
can be derived as

\begin{equation}
\mathcal{E}_{n}\left\{ e^{\frac{-\left|\mathbf{h}_{k}\mathbf{W}\mathbf{x}_{m,i}+n_{k}\right|^{2}}{\sigma_{k}^{2}}}\right\} =e^{-\frac{\left|\mathbf{h}_{k}\mathbf{W}\mathbf{x}_{m,i}\right|^{2}}{2\sigma_{k}^{2}}}.\label{eq:12-1-1}
\end{equation}

Now, we can write $\varphi$ as 

\[
\varphi=\log_{2}\stackrel[i=1]{M^{N}}{\sum}\mathcal{E}_{\mathbf{h}}\left\{ e^{-\frac{\left|\mathbf{h}_{k}\mathbf{W}\mathbf{x}_{m,i}\right|^{2}}{2\sigma_{k}^{2}}}\right\} ,\quad\quad\quad\quad\quad\quad
\]

\[
=\log_{2}\stackrel[i=1]{M^{N}}{\sum}\mathcal{E}_{\mathbf{h}}\left\{ e^{-\frac{\left|\sqrt{P_{c}}\mathbf{h}_{k}\mathbf{w}_{c}\left[\mathbf{x}_{m,i}\right]_{1}+\sqrt{P_{p}}\mathbf{h}_{k}\mathbf{W}_{ZF}^{p}\mathbf{x}_{m,i}^{p}\right|^{2}}{2\sigma_{k}^{2}}}\right\} ,
\]

\begin{equation}
=\log_{2}\stackrel[i=1]{M^{N}}{\sum}\mathcal{E}_{\mathbf{h}}\left\{ e^{-\frac{P\left|\sqrt{\left(1-t\right)}\beta_{c}\left(\overset{K}{\underset{i=1}{\sum}}\mathbf{h}_{k}\mathbf{h}_{i}^{H}\right)\left[\mathbf{x}_{m,i}\right]_{1}+\sqrt{t}\beta_{p}\left[\mathbf{x}_{m,i}^{p}\right]_{k}\right|^{2}}{2\sigma_{k}^{2}}}\right\} .
\end{equation}

\noindent Since the term $Y=\left(\overset{K}{\underset{i=1}{\sum}}\mathbf{h}_{k}\mathbf{h}_{i}^{H}\right)$
has Gamma distribution, .i.e., $Y\thicksim\Gamma\left(\upsilon,\theta\right)$
,  the average can be derived as

\begin{equation}
\varphi=\log_{2}\stackrel[i=1]{M^{N}}{\sum}\stackrel[0]{\infty}{\int}e^{-\frac{P\left|\sqrt{\left(1-t\right)}\beta_{c}y\left[\mathbf{x}_{m,i}\right]_{1}+\sqrt{t}\beta_{p}\left[\mathbf{x}_{m,i}^{p}\right]_{k}\right|^{2}}{2\sigma_{k}^{2}}}\frac{y^{\upsilon-1}e^{\frac{-y}{\theta}}}{\Gamma\left(\upsilon\right)\theta^{\upsilon}}\, dy.
\end{equation}
Applying Gaussian Quadrature rule, the average can be obtained by,

\begin{equation}
\varphi=\log_{2}\stackrel[i=1]{M^{N}}{\sum}\stackrel[r=1]{n}{\sum}\frac{\left(y_{r}\right)^{\upsilon-1}\textrm{H}_{r}}{\Gamma\left(\upsilon\right)}e^{-\frac{P\left|\sqrt{\left(1-t\right)}\beta_{c}\theta y_{r}\left[\mathbf{x}_{m,i}\right]_{1}+\sqrt{t}\beta_{p}\left[\mathbf{x}_{m,i}^{p}\right]_{k}\right|^{2}}{2\sigma_{k}^{2}}}
\end{equation}

\noindent where $y_{r}$ and $\textrm{H}_{r}$ are the $r^{th}$ zero
and the weighting factor of the Laguerre polynomials, respectively
\cite{book}. Similarly, for the second term $\psi$, using Jensen
inequality we can write, 

\begin{equation}
\psi=\log_{2}\stackrel[t=1]{M^{N-1}}{\sum}\mathcal{E}_{n_{k}}\left\{ e^{-\frac{\left|\sqrt{tP}\beta_{p}\left[\mathbf{x}_{m,i}^{p}\right]_{k}+n_{k}\right|^{2}}{\sigma_{k}^{2}}}\right\} .
\end{equation}

The average over the noise can be obtained as

\begin{equation}
\psi=\log_{2}\stackrel[t=1]{M^{N-1}}{\sum}e^{-\frac{\left|\sqrt{tP}\beta_{p}\left[\mathbf{x}_{m,i}^{p}\right]_{k}\right|^{2}}{2\sigma_{k}^{2}}}.\label{eq:39}
\end{equation}

\subsubsection{\noindent Ergodic Rate for the Private Part}

\noindent The ergodic rate for the private message at the $k^{th}$
user, under PSK signaling using ZF precoding technique can be written
as\cite{interference1,precoding6,precoding2},

\begin{equation}
\mathcal{E}\left\{ R_{k}^{p}\right\} =\log_{2}M-\,\frac{1}{M^{N-1}}\stackrel[m=1]{M^{N-1}}{\sum}\underset{\psi}{\underbrace{\mathcal{E}\left\{ \log_{2}\stackrel[i=1]{M^{N-1}}{\sum}e^{\frac{-\left|\mathbf{h}_{k}\mathbf{W}_{ZF}^{p}\mathbf{x}_{m,i}^{p}\right|^{2}}{2\sigma_{k}^{2}}}\right\} }}.\label{eq:38}
\end{equation}

\noindent By using Jensen inequality, and following similar steps
as in the previous sub-section we can find the average of the term
$\psi$ in (\ref{eq:38}) as in (\ref{eq:39}).

\subsection{Conventional Transmission Without Rate Splitting (NoRS)}

The ergodic rate at the $k^{th}$ user in conventional transmission
without RS is expressed by

\begin{equation}
\mathcal{E}\left\{ R_{k}^{NoRS}\right\} =\log_{2}M-\frac{1}{M^{N}}\stackrel[m=1]{M^{N}}{\sum}\underset{\psi}{\underbrace{\mathcal{E}\left\{ \log_{2}\stackrel[i=1]{M^{N}}{\sum}e^{\frac{-\left|\mathbf{h}_{k}\mathbf{W}\mathbf{x}_{m,i}\right|^{2}}{2\sigma_{k}^{2}}}\right\} }}.\label{eq:35}
\end{equation}

In CI case, the precoding matrix $\mathbf{W}$ is given in (\ref{eq:7}),
and the expectation in (\ref{eq:35}) can be derived using Jensen
inequality as in (\ref{eq:23}). On the other hand, in ZF scenario
the precoding matrix $\mathbf{W}$ is given in (\ref{eq:24}), and
then the expectation in (\ref{eq:35}) can be derived using Jensen
inequality as in (\ref{eq:39}).

Please note that, in case the users' locations are randomly distributed,
the ergodic sum-rate with respect to each user location can be calculated
easily by averaging the derived sum-rate expression over all possible
user locations.

\section{Ergodic Sum Rate Analysis under PSK signaling and Imperfect CSI}

In practice, the BS can estimate the channel matrix \textbf{$\boldsymbol{\textrm{H}}$}
by transmitting pilot signals. Therefore, the current channels in
terms of the estimated channels, and the estimation error can be written
as \cite{marzita,Rs4}, $\textrm{\ensuremath{\boldsymbol{H}}}=\hat{\textrm{\ensuremath{\boldsymbol{H}}}}+\textrm{E}$,
where $\hat{\textrm{\ensuremath{\boldsymbol{H}}}}$ is the estimated
channel matrix, $\textrm{E}$ is the estimation error matrix. The
two matrices $\hat{\textrm{\ensuremath{\boldsymbol{H}}}}$, and $\textrm{E}$
are assumed to be mutually independent and distributed as $\hat{\textrm{\ensuremath{\boldsymbol{H}}}}\sim\mathcal{CN}\left(\text{0,}\hat{\textrm{\ensuremath{\boldsymbol{D}}}}\right)$
and $\textrm{E}\sim\mathcal{CN}\left(\text{0,}\textrm{\ensuremath{\boldsymbol{D}}}-\hat{\textrm{\ensuremath{\boldsymbol{D}}}}\right)$,
where $\hat{\textrm{\ensuremath{\boldsymbol{D}}}}$ is a diagonal
matrix with $\left[\hat{\mathbf{D}}\right]_{kk}=\hat{\sigma}_{k}^{2}=\frac{p_{u}\varpi_{k}^{2}}{p_{u}\varpi_{k}+1}$
and $\left[\textrm{\ensuremath{\boldsymbol{D}}}-\hat{\mathbf{D}}\right]_{kk}=\hat{\sigma}_{ek}^{2}=\frac{\varpi_{k}}{p_{u}\varpi_{k}+1}$
\cite{marzita,Rs4}, while $p_{u}=\tau p_{p}$ and $\varpi_{k}=d_{k}^{-m}$,
$\tau$ is number of symbols used for channel training and $p_{p}$
is the transmit power for each pilot symbol. Consequently, the received
signal can be written now as,

\begin{equation}
\hat{y}_{k}=\sqrt{P_{c}}\mathbf{\hat{h}}_{k}\mathbf{\hat{w}}_{c}x_{c}-\sqrt{P_{c}}\boldsymbol{\boldsymbol{e}}_{k}\mathbf{\hat{w}}_{c}x_{c}+\stackrel[i=1]{K}{\sum}\sqrt{P_{p}}\mathbf{\hat{h}}_{k}\mathbf{\hat{w}}_{i}^{p}x_{i}-\stackrel[i=1]{K}{\sum}\sqrt{P_{p}}\boldsymbol{\boldsymbol{e}}_{k}\mathbf{\hat{w}}_{i}^{p}x_{i}+n_{k}.
\end{equation}

\subsection{RS: MRT/CI}

In this scenario, the precoder for the common and the private messages
based on the estimated channels can be written, respectively, as \cite{Rs6,angLi}

\begin{equation}
\mathbf{\hat{w}}_{c}=\overset{K}{\underset{i=1}{\sum}}\beta_{c}\mathbf{\hat{h}}_{i}^{H}
\end{equation}

\begin{equation}
\mathbf{\hat{W}}^{p}=\frac{1}{K}\beta_{p}\,\mathbf{\hat{H}}^{H}\left(\mathbf{\hat{H}}\mathbf{\hat{H}}^{H}\right)^{-1}\textrm{diag}\left\{ \mathbf{\hat{V}}^{-1}\mathbf{u}\right\} .\label{eq:7-2}
\end{equation}

\noindent The received signal at user $k$ can be now written as 

\begin{equation}
\hat{y}_{k}=\sqrt{P_{c}}\beta_{c}\overset{K}{\underset{i=1}{\sum}}\mathbf{\hat{h}}_{k}\mathbf{\hat{h}}_{i}^{H}x_{c}-\sqrt{P_{c}}\beta_{c}\overset{K}{\underset{i=1}{\sum}}\boldsymbol{\boldsymbol{e}}_{k}\mathbf{\hat{h}}_{i}^{H}x_{c}+\sqrt{P_{p}}\stackrel[i=1]{K}{\sum}\mathbf{\hat{h}}_{k}\mathbf{\hat{w}}_{i}^{p}x_{i}-\sqrt{P_{p}}\stackrel[i=1]{K}{\sum}\boldsymbol{\boldsymbol{e}}_{k}\mathbf{\hat{w}}_{i}^{p}x_{i}+n_{k}.
\end{equation}

\subsubsection{\noindent Ergodic Rate for the Common Part}

\noindent The ergodic rate for the common part at user $k$ under
PSK signaling in imperfect CSIT scenario, can be written as\cite{interference1,precoding6,precoding2}

\[
\mathcal{E}\left\{ R_{k}^{c}\right\} =\log_{2}M-\frac{1}{M^{N}}\stackrel[m=1]{M^{N}}{\sum}\underset{\varphi}{\underbrace{\mathcal{E}_{\mathbf{\hat{h}},\boldsymbol{\boldsymbol{e}},n_{k}}\left\{ \log_{2}\stackrel[i=1]{M^{N}}{\sum}e^{\frac{-\left|\mathbf{\hat{h}}_{k}\mathbf{\hat{W}}\mathbf{x}_{m,i}+\boldsymbol{\boldsymbol{e}}_{k}\mathbf{\hat{W}}\mathbf{x}_{m,i}+n_{k}\right|^{2}}{\sigma_{k}^{2}}}\right\} }}
\]

\begin{equation}
+\frac{1}{M^{N-1}}\stackrel[m=1]{M^{N-1}}{\sum}\underset{\psi}{\underbrace{\mathcal{E}_{\mathbf{\hat{h}},\boldsymbol{\boldsymbol{e}},n_{k}}\left\{ \log_{2}\stackrel[i=1]{M^{N-1}}{\sum}e^{\frac{-\left|\mathbf{\hat{h}}_{k}\mathbf{\hat{W}}^{p}\mathbf{x}_{m,i}^{p}+\boldsymbol{\boldsymbol{e}}_{k}\mathbf{\hat{W}}\mathbf{x}_{m,i}+n_{k}\right|^{2}}{\sigma_{k}^{2}}}\right\} }}.\label{eq:7-1-2}
\end{equation}

As one can see from (\ref{eq:7-1-2}), the ergodic rate is hard to
further simplify, since the expectations involve several random variables.
However, an approximation based on large number of antennas at the
BS can be derived.

\noindent \textbf{\textit{Analysis for Large $N$}}

\noindent In this case we analyze the ergodic rate when the number
of BS antennas is large $\left(N\gg K\right)$,  driven by the increasing
research interest in MU-MIMO systems with a large number of BS antennas.

\begin{lem}
Let $\boldsymbol{a}=\left[a_{1}.....a_{n}\right]^{T}$ and $\boldsymbol{b}=\left[b_{1}.....b_{n}\right]^{T}$
be $n\times1$ independent vectors contain i.i.d entries with zero-mean
and variances $\mathcal{E}\left\{ \left|a_{i}\right|^{2}\right\} =\sigma_{a}^{2}$
and $\mathcal{E}\left\{ \left|b_{i}\right|^{2}\right\} =\sigma_{b}^{2}$.
Therefore, following the law of large numbers, we can get \cite{marzita}

\begin{equation}
\frac{1}{n}\boldsymbol{a}^{H}\boldsymbol{a}\stackrel{\textrm{a.s}}{\rightarrow}\sigma_{a}^{2}\textrm{,}\,\frac{1}{n}\boldsymbol{b}^{H}\boldsymbol{b}\stackrel{\textrm{a.s}}{\rightarrow}\sigma_{b}^{2}\textrm{ and }\frac{1}{n}\boldsymbol{a}^{H}\boldsymbol{b}\stackrel{\textrm{a.s}}{\rightarrow}0,\,\,\label{eq:57}
\end{equation}

\begin{equation}
\frac{1}{\sqrt{n}}\boldsymbol{a}^{H}\boldsymbol{b}\stackrel{\textrm{d}}{\rightarrow}\mathcal{CN}\left(\text{0,}\sigma_{a}^{2}\sigma_{b}^{2}\right),\,\,\label{eq:58}
\end{equation}

\noindent where $\stackrel{\textrm{a.s}}{\rightarrow}$ and $\stackrel{\textrm{d}}{\rightarrow}$denote
almost-sure and distribution convergence, respectively.
\end{lem}
It is well known that by deploying very large number of antennas at
the BS, the small-scale fading can be averaged out. Therefore, we
now can elaborate more on analyzing the impact of large-scale fading
on the system performance. Using the facts in Lemma 1, (\ref{eq:7-1-2})
becomes

\[
\mathcal{E}\left\{ R_{k}^{c}\right\} =\log_{2}M-\frac{1}{M^{N}}\stackrel[m=1]{M^{N}}{\sum}\underset{\varphi}{\underbrace{\mathcal{E}_{d_{k},n_{k}}\left\{ \log_{2}\stackrel[i=1]{M^{N}}{\sum}e^{\frac{-\left|\sqrt{P_{c}}\beta_{c}N\hat{\sigma}_{k}^{2}x_{m,i}^{c}+\sqrt{P_{p}}N\frac{1}{K}\beta_{p}u_{k}\hat{\sigma}_{k}^{2}x_{m,i}^{p}+n_{k}\right|^{2}}{\sigma_{k}^{2}}}\right\} }}
\]

\begin{equation}
+\frac{1}{M^{N-1}}\stackrel[m=1]{M^{N-1}}{\sum}\underset{\psi}{\underbrace{\mathcal{E}_{d_{k},n_{k}}\left\{ \log_{2}\stackrel[i=1]{M^{N-1}}{\sum}e^{\frac{-\left|\sqrt{P_{p}}N\frac{1}{K}\beta_{p}u_{k}\hat{\sigma}_{k}^{2}x_{m,i}^{p}+n_{k}\right|^{2}}{\sigma_{k}^{2}}}\right\} }},
\end{equation}

and 

\[
\mathcal{E}\left\{ R_{k}^{c}\right\} =\log_{2}M-\frac{1}{M^{N}}\stackrel[m=1]{M^{N}}{\sum}\underset{\varphi}{\underbrace{\mathcal{E}_{d_{k},n_{k}}\left\{ \log_{2}\stackrel[i=1]{M^{N}}{\sum}e^{\frac{-\left|\sqrt{P_{c}}\beta_{c}N\left(\frac{p_{u}\varpi_{k}^{2}}{p_{u}\varpi_{k}+1}\right)x_{m,i}^{c}+\sqrt{P_{p}}N\frac{1}{K}\beta_{p}u_{k}\left(\frac{p_{u}\varpi_{k}^{2}}{p_{u}\varpi_{k}+1}\right)x_{m,i}^{p}+n_{k}\right|^{2}}{\sigma_{k}^{2}}}\right\} }}
\]

\begin{equation}
+\frac{1}{M^{N-1}}\stackrel[m=1]{M^{N-1}}{\sum}\underset{\psi}{\underbrace{\mathcal{E}_{d_{k},n_{k}}\left\{ \log_{2}\stackrel[i=1]{M^{N-1}}{\sum}e^{\frac{-\left|\sqrt{P_{p}}N\frac{1}{K}\beta_{p}u_{k}\left(\frac{p_{u}\varpi_{k}^{2}}{p_{u}\varpi_{k}+1}\right)x_{m,i}^{p}+n_{k}\right|^{2}}{\sigma_{k}^{2}}}\right\} }}.\label{eq:65}
\end{equation}

\noindent By invoking Jensen inequality, the first term in (\ref{eq:65}),
$\varphi$, can be expressed by

\[
\varphi=\mathcal{E}_{d_{k},n_{k}}\left\{ \log_{2}\stackrel[i=1]{M^{N}}{\sum}e^{\frac{-\left|\sqrt{P_{c}}\beta_{c}N\left(\frac{p_{u}\varpi_{k}^{2}}{p_{u}\varpi_{k}+1}\right)x_{m,i}^{c}+\sqrt{P_{p}}N\frac{1}{K}\beta_{p}u_{k}\left(\frac{p_{u}\varpi_{k}^{2}}{p_{u}\varpi_{k}+1}\right)x_{m,i}^{p}+n_{k}\right|^{2}}{\sigma_{k}^{2}}}\right\} \leq
\]

\begin{equation}
\quad\quad\quad\quad\quad\log_{2}\stackrel[i=1]{M^{N}}{\sum}\mathcal{E}_{d_{k},n_{k}}\left\{ e^{\frac{-\left|\sqrt{P_{c}}\beta_{c}N\left(\frac{p_{u}\varpi_{k}^{2}}{p_{u}\varpi_{k}+1}\right)x_{m,i}^{c}+\sqrt{P_{p}}N\frac{1}{K}\beta_{p}u_{k}\left(\frac{p_{u}\varpi_{k}^{2}}{p_{u}\varpi_{k}+1}\right)x_{m,i}^{p}+n_{k}\right|^{2}}{\sigma_{k}^{2}}}\right\} .
\end{equation}

Since the noise $n_{k}$ has Gaussian distribution, using the integrals
of exponential function, we can find \cite{book2}

\[
\mathcal{E}_{n}\left\{ e^{\frac{-\left|\sqrt{P_{c}}\beta_{c}N\left(\frac{p_{u}\varpi_{k}^{2}}{p_{u}\varpi_{k}+1}\right)x_{m,i}^{c}+\sqrt{P_{p}}N\frac{1}{K}\beta_{p}u_{k}\left(\frac{p_{u}\varpi_{k}^{2}}{p_{u}\varpi_{k}+1}\right)x_{m,i}^{p}+n_{k}\right|^{2}}{\sigma_{k}^{2}}}\right\} =
\]

\begin{equation}
\qquad\qquad\qquad e^{-\frac{\left|\sqrt{P_{c}}\beta_{c}N\left(\frac{p_{u}\varpi_{k}^{2}}{p_{u}\varpi_{k}+1}\right)x_{m,i}^{c}+\sqrt{P_{p}}N\frac{1}{K}\beta_{p}u_{k}\left(\frac{p_{u}\varpi_{k}^{2}}{p_{u}\varpi_{k}+1}\right)x_{m,i}^{p}\right|^{2}}{2\sigma_{k}^{2}}}.\label{eq:12-1-2}
\end{equation}

Now, the average over the user location can be derived as

\[
\mathcal{E}_{d_{k}}\left\{ e^{-\frac{\left|\sqrt{P_{c}}\beta_{c}N\left(\frac{p_{u}\varpi_{k}^{2}}{p_{u}\varpi_{k}+1}\right)x_{m,i}^{c}+\sqrt{P_{p}}N\frac{1}{K}\beta_{p}u_{k}\left(\frac{p_{u}\varpi_{k}^{2}}{p_{u}\varpi_{k}+1}\right)x_{m,i}^{p}\right|^{2}}{2\sigma_{k}^{2}}}\right\} =
\]

\begin{equation}
\quad\quad\quad\quad\quad\mathcal{E}_{d_{k}}\left\{ e^{-\frac{\left|\frac{p_{u}}{p_{u}d_{k}^{m}+d_{k}^{2m}}\right|^{2}\left|\sqrt{P_{c}}\beta_{c}Nx_{m,i}^{c}+\sqrt{P_{p}}N\frac{1}{K}\beta_{p}u_{k}x_{m,i}^{p}\right|^{2}}{2\sigma_{k}^{2}}}\right\} .
\end{equation}

For analytical convenience, in this section we assume that the cell
shape is approximated by a circle of radius $R$, and the users are
uniformly distributed in the cell \cite{salim}. Hence, the PDF of
the users at radius $r$ relative to the BS is \cite{salim} $f_{d}\left(r\right)=\frac{2\left(r-R_{0}\right)}{\left(R-R_{0}\right)^{2}},\quad R_{0}\leq r\leq R$,
where $R_{0}$ is the closest distance between a user and the BS.
Therefore, we can find the average over $d_{k}$ using Gaussian Quadrature
rules as, 

\begin{equation}
\mathcal{E}_{d_{k}}\left\{ e^{-\frac{\left|\frac{p_{u}}{p_{u}d_{k}^{m}+d_{k}^{2m}}\right|^{2}\zeta}{2\sigma_{k}^{2}}}\right\} =\stackrel[R_{0}]{R}{\int}e^{-\frac{\left|\frac{p_{u}}{p_{u}r^{m}+r^{2m}}\right|^{2}\zeta}{2\sigma_{k}^{2}}}\frac{2\left(r-R_{0}\right)}{\left(R-R_{0}\right)^{2}}dr\label{eq:71}
\end{equation}

\begin{equation}
=\stackrel[j=1]{C}{\sum}\textrm{H}_{j}\, e^{-\frac{\left|\frac{p_{u}}{p_{u}\left(\frac{R-R_{0}}{2}\, r_{j}+\frac{R+R_{0}}{2}\right)^{m}+\left(\frac{R-R_{0}}{2}\, r_{j}+\frac{R+R_{0}}{2}\right)^{2m}}\right|^{2}\zeta}{2\sigma_{k}^{2}}}\frac{2\left(\left(\frac{R-R_{0}}{2}\, r_{j}+\frac{R+R_{0}}{2}\right)-R_{0}\right)}{\left(R-R_{0}\right)^{2}}\label{eq:72}
\end{equation}

\noindent where $\zeta=\left|\sqrt{P_{c}}\beta_{c}Nx_{m,i}^{c}+\sqrt{P_{p}}N\frac{1}{K}\beta_{p}u_{k}x_{m,i}^{p}\right|^{2}$and
$r_{j}$ and $\textrm{H}_{j}$ are the $j^{th}$ zero and the weighting
factors of the Laguerre polynomials, respectively \cite{book}.

\noindent For the second term, $\psi$, similarly using Jensen inequality
we can write

\[
\psi=\mathcal{E}_{d_{k},n_{k}}\left\{ \log_{2}\stackrel[i=1]{M^{N-1}}{\sum}e^{\frac{-\left|\sqrt{P_{p}}N\frac{1}{K}\beta_{p}u_{k}\left(\frac{p_{u}\varpi_{k}^{2}}{p_{u}\varpi_{k}+1}\right)x_{m,i}^{p}+n_{k}\right|^{2}}{\sigma_{k}^{2}}}\right\} \leq
\]

\begin{equation}
\quad\quad\quad\quad\log_{2}\stackrel[i=1]{M^{N-1}}{\sum}\mathcal{E}_{d_{k},n_{k}}\left\{ e^{\frac{-\left|\sqrt{P_{p}}N\frac{1}{K}\beta_{p}u_{k}\left(\frac{p_{u}\varpi_{k}^{2}}{p_{u}\varpi_{k}+1}\right)x_{m,i}^{p}+n_{k}\right|^{2}}{\sigma_{k}^{2}}}\right\} .
\end{equation}

\noindent Since $n_{k}$ has Gaussian distribution, we can get 

\begin{equation}
\psi=\log_{2}\stackrel[i=1]{M^{N-1}}{\sum}\mathcal{E}_{d_{k}}\left\{ e^{-\frac{\left|\frac{p_{u}}{p_{u}d_{k}^{m}+d_{k}^{2m}}\right|^{2}\vartheta}{2\sigma_{k}^{2}}}\right\} \label{eq:18-2}
\end{equation}

\noindent where $\vartheta=\left|\sqrt{P_{p}}N\frac{1}{K}\beta_{p}u_{k}x_{m,i}^{p}\right|^{2}$.
The average in (\ref{eq:18-2}) can be obtained as in (\ref{eq:71})
and (\ref{eq:72}), which is given by 

\begin{equation}
\mathcal{E}_{d_{k}}\left\{ e^{-\frac{\left|\frac{p_{u}}{p_{u}d_{k}^{m}+d_{k}^{2m}}\right|^{2}\vartheta}{2\sigma_{k}^{2}}}\right\} =\stackrel[R_{0}]{R}{\int}e^{-\frac{\left|\frac{p_{u}}{p_{u}r^{m}+r^{2m}}\right|^{2}\vartheta}{2\sigma_{k}^{2}}}\frac{2\left(r-R_{0}\right)}{\left(R-R_{0}\right)^{2}}dr\label{eq:71-1}
\end{equation}

\begin{equation}
=\stackrel[j=1]{C}{\sum}\textrm{H}_{j}\, e^{-\frac{\left|\frac{p_{u}}{p_{u}\left(\frac{R-R_{0}}{2}\, r_{j}+\frac{R+R_{0}}{2}\right)^{m}+\left(\frac{R-R_{0}}{2}\, r_{j}+\frac{R+R_{0}}{2}\right)^{2m}}\right|^{2}\vartheta}{2\sigma_{k}^{2}}}\frac{2\left(\left(\frac{R-R_{0}}{2}\, r_{j}+\frac{R+R_{0}}{2}\right)-R_{0}\right)}{\left(R-R_{0}\right)^{2}}.\label{eq:72-1}
\end{equation}

\subsubsection{\noindent Ergodic Rate for the Private Part}

\noindent The ergodic rate for the private part at user $k$ under
PSK signaling, using CI precoding technique can be written as\cite{interference1,precoding6,precoding2},

\begin{equation}
\mathcal{E}\left\{ R_{k}^{p}\right\} =\log_{2}M-\frac{1}{M^{N-1}}\stackrel[m=1]{M^{N-1}}{\sum}\underset{\psi}{\underbrace{\mathcal{E}_{\mathbf{h},n_{k}}\left\{ \log_{2}\stackrel[i=1]{M^{N-1}}{\sum}e^{\frac{-\left|\mathbf{h}_{k}\mathbf{W}_{CI}^{p}\mathbf{x}_{m,i}^{p}\right|^{2}}{2\sigma_{k}^{2}}}\right\} }}.\label{eq:65-2}
\end{equation}

\noindent By using Jensen inequality, and following similar steps
as in the previous sub-section we can find the average of $\psi$
in (\ref{eq:65-2}) as in (\ref{eq:18-2}) and (\ref{eq:72-1}).

\subsection{RS: MRT/ZF}

In this case the precoding for the common and the private messages
based on the estimated channels can be written, respectively, as

\begin{equation}
\mathbf{\hat{w}}_{c}=\overset{K}{\underset{i=1}{\sum}}\beta_{c}\mathbf{\hat{h}}_{i}^{H}
\end{equation}

\begin{equation}
\mathbf{\hat{W}}^{p}=\beta_{p}\,\mathbf{\hat{H}}^{H}\left(\mathbf{\hat{H}}\mathbf{\hat{H}}^{H}\right)^{-1}.\label{eq:24-1}
\end{equation}

\noindent Therefore, the received signal is given by

\[
\hat{y}_{k}=\sqrt{P_{c}}\beta_{c}N\overset{K}{\underset{i=1}{\sum}}\frac{1}{N}\mathbf{\hat{h}}_{k}\mathbf{\hat{h}}_{i}^{H}x_{c}-\sqrt{P_{c}}\beta_{c}N\overset{K}{\underset{i=1}{\sum}}\frac{1}{N}\boldsymbol{\boldsymbol{e}}_{k}\mathbf{\hat{h}}_{i}^{H}x_{c}
\]

\begin{equation}
\qquad\qquad\qquad+\sqrt{P_{p}}\stackrel[i=1]{K}{\sum}\mathbf{\hat{h}}_{k}\mathbf{\hat{w}}_{i}^{p}x_{i}-\sqrt{P_{p}}N\stackrel[i=1]{K}{\sum}\frac{1}{N}\boldsymbol{\boldsymbol{e}}_{k}\mathbf{\hat{w}}_{i}^{p}x_{i}+n_{k}.
\end{equation}

\subsubsection{\noindent Ergodic Rate for the Common Part}

\noindent The ergodic rate for the common part at user $k$ under
PSK signaling in imperfect CSI scenario can be written as\cite{interference1,precoding6,precoding2},

\[
\mathcal{E}\left\{ R_{k}^{c}\right\} =\log_{2}M-\frac{1}{M^{N}}\stackrel[m=1]{M^{N}}{\sum}\underset{\varphi}{\underbrace{\mathcal{E}_{\mathbf{\hat{h}},\boldsymbol{\boldsymbol{e}},n_{k}}\left\{ \log_{2}\stackrel[i=1]{M^{N}}{\sum}e^{\frac{-\left|\mathbf{\hat{h}}_{k}\mathbf{\hat{W}}\mathbf{x}_{m,i}+\boldsymbol{\boldsymbol{e}}_{k}\mathbf{\hat{W}}\mathbf{x}_{m,i}+n_{k}\right|^{2}}{\sigma_{k}^{2}}}\right\} }}
\]

\begin{equation}
+\frac{1}{M^{N-1}}\stackrel[m=1]{M^{N-1}}{\sum}\underset{\psi}{\underbrace{\mathcal{E}_{\mathbf{\hat{h}},\boldsymbol{\boldsymbol{e}},n_{k}}\left\{ \log_{2}\stackrel[i=1]{M^{N-1}}{\sum}e^{\frac{-\left|\mathbf{\hat{h}}_{k}\mathbf{\hat{W}}^{p}\mathbf{x}_{m,i}^{p}+\boldsymbol{\boldsymbol{e}}_{k}\mathbf{\hat{W}}\mathbf{x}_{m,i}+n_{k}\right|^{2}}{\sigma_{k}^{2}}}\right\} }}.\label{eq:7-1-2-1}
\end{equation}

For the sake of comparison, here we derive an approximation of the
user rate based on a large number of antennas.

\noindent \textbf{\textit{Analysis for Large $N$}}

\noindent The rate for the common part at user $k$ when $\left(N\gg K\right)$
can be written as

\[
\mathcal{E}\left\{ R_{k}^{c}\right\} =N\,\log_{2}M-\frac{1}{M^{N}}\stackrel[m=1]{M^{N}}{\sum}\underset{\varphi}{\underbrace{\mathcal{E}_{d_{k},n_{k}}\left\{ \log_{2}\stackrel[i=1]{M^{N}}{\sum}e^{\frac{-\left|\sqrt{P_{c}}\beta_{c}N\hat{\sigma}_{k}^{2}x_{m,i}^{c}+\sqrt{P_{p}}\beta_{p}x_{m,i}^{p}+n_{k}\right|^{2}}{\sigma_{k}^{2}}}\right\} }}
\]

\begin{equation}
+\frac{1}{M^{N-1}}\stackrel[m=1]{M^{N-1}}{\sum}\underset{\psi}{\underbrace{\mathcal{E}_{d_{k},n_{k}}\left\{ \log_{2}\stackrel[i=1]{M^{N-1}}{\sum}e^{\frac{-\left|\sqrt{P_{p}}\beta_{p}x_{m,i}^{p}+n_{k}\right|^{2}}{\sigma_{k}^{2}}}\right\} }}.\label{eq:7-1-1-1}
\end{equation}
By using Jensen inequality, the first term in (\ref{eq:7-1-1-1}),
$\varphi$, can be expressed by

\[
\varphi=\mathcal{E}_{d_{k},n_{k}}\left\{ \log_{2}\stackrel[i=1]{M^{N}}{\sum}e^{\frac{-\left|\sqrt{P_{c}}\beta_{c}N\hat{\sigma}_{k}^{2}x_{m,i}^{c}+\sqrt{P_{p}}\beta_{p}x_{m,i}^{p}+n_{k}\right|^{2}}{\sigma_{k}^{2}}}\right\} \leq
\]

\begin{equation}
\log_{2}\stackrel[i=1]{M^{N}}{\sum}\mathcal{E}_{d_{k},n_{k}}\left\{ e^{\frac{-\left|\sqrt{P_{c}}\beta_{c}N\hat{\sigma}_{k}^{2}x_{m,i}^{c}+\sqrt{P_{p}}\beta_{p}x_{m,i}^{p}+n_{k}\right|^{2}}{\sigma_{k}^{2}}}\right\} .
\end{equation}

Since the noise $n_{k}$ has Gaussian distribution, the average over
the noise using the integrals of exponential function can be derived
as \cite{book2} 

\begin{equation}
\mathcal{E}_{n}\left\{ e^{\frac{-\left|\sqrt{P_{c}}\beta_{c}N\hat{\sigma}_{k}^{2}x_{m,i}^{c}+\sqrt{P_{p}}\beta_{p}N\hat{\sigma}_{k}^{2}x_{m,i}^{p}+n_{k}\right|^{2}}{\sigma_{k}^{2}}}\right\} =e^{-\frac{\left|\sqrt{P_{c}}\beta_{c}N\hat{\sigma}_{k}^{2}x_{m,i}^{c}+\sqrt{P_{p}}\beta_{p}x_{m,i}^{p}\right|^{2}}{2\sigma_{k}^{2}}}.
\end{equation}

Now, we can write $\varphi$ as 

\begin{equation}
\varphi=\log_{2}\stackrel[i=1]{M^{N}}{\sum}\mathcal{E}_{d_{k}}\left\{ e^{-\frac{\left|\sqrt{P_{c}}\beta_{c}N\hat{\sigma}_{k}^{2}x_{m,i}^{c}+\sqrt{P_{p}}\beta_{p}x_{m,i}^{p}\right|^{2}}{2\sigma_{k}^{2}}}\right\} .
\end{equation}

Similarly to the CI scenario, we assume that the cell shape is approximated
by a circle of radius $R$ and the users are uniformly distributed
in the cell \cite{salim}. Therefore, we can find the average over
$d_{k}$ by 

\begin{equation}
\mathcal{E}_{d_{k}}\left\{ e^{-\frac{\left|\sqrt{P_{c}}\beta_{c}N\hat{\sigma}_{k}^{2}x_{c}+\sqrt{P_{p}}\beta_{p}x_{k}\right|^{2}}{2\sigma_{k}^{2}}}\right\} =\stackrel[R_{0}]{R}{\int}e^{-\frac{\left|\sqrt{P_{c}}\beta_{c}N\hat{\sigma}_{k}^{2}x_{m,i}^{c}+\sqrt{P_{p}}\beta_{p}x_{m,i}^{p}\right|^{2}}{2\sigma_{k}^{2}}}\frac{2\left(r-R_{0}\right)}{\left(R-R_{0}\right)^{2}}dr
\end{equation}

\noindent  which can be found using Gaussian Quadrature rules as

\[
\stackrel[R_{0}]{R}{\int}e^{-\frac{\left|\sqrt{P_{c}}\beta_{c}N\left(\frac{\tau p_{p}\varpi_{k}^{2}}{\tau p_{p}\varpi_{k}+1}\right)x_{m,i}^{c}+\sqrt{P_{p}}\beta_{p}x_{m,i}^{p}\right|^{2}}{2\sigma_{k}^{2}}}\frac{2\left(r-R_{0}\right)}{\left(R-R_{0}\right)^{2}}dr=
\]

\begin{equation}
\stackrel[j=1]{C}{\sum}\textrm{H}_{j}\, e^{-\frac{\left|\sqrt{P_{c}}\beta_{c}N\left(\frac{p_{u}}{p_{u}\left(\frac{R-R_{0}}{2}\, r_{j}+\frac{R+R_{0}}{2}\right)^{m}+\left(\frac{R-R_{0}}{2}\, r_{j}+\frac{R+R_{0}}{2}\right)^{2m}}\right)x_{m,i}^{c}+\sqrt{P_{p}}\beta_{p}x_{m,i}^{p}\right|^{2}}{2\sigma_{k}^{2}}}\frac{2\left(\left(\frac{R-R_{0}}{2}\, r_{j}+\frac{R+R_{0}}{2}\right)-R_{0}\right)}{\left(R-R_{0}\right)^{2}}.
\end{equation}

\noindent For the second term $\psi$, using Jensen inequality we
can write

\begin{equation}
\psi=\log_{2}\stackrel[t=1]{M^{N-1}}{\sum}\mathcal{E}_{n_{k}}\left\{ e^{-\frac{\left|\sqrt{P_{p}}\beta_{p}x_{m,i}^{p}+n_{k}\right|^{2}}{\sigma_{k}^{2}}}\right\} .
\end{equation}

Since the noise $n_{k}$ has Gaussian distribution, the average can
be derived as

\begin{equation}
\psi=\log_{2}\stackrel[t=1]{M^{N-1}}{\sum}e^{-\frac{\left|\sqrt{tP}\beta_{p}x_{m,i}^{p}\right|^{2}}{2\sigma_{k}^{2}}}.\label{eq:39-2}
\end{equation}

\subsubsection{\noindent Ergodic Rate for the Private Part}

\noindent The ergodic rate for the private message at the $k^{th}$
user, under PSK signaling using ZF precoding technique can be written
as\cite{interference1,precoding6,precoding2}

\begin{equation}
\mathcal{E}\left\{ R_{k}^{p}\right\} =\log_{2}M-\frac{1}{M^{N-1}}\stackrel[m=1]{M^{N-1}}{\sum}\underset{\psi}{\underbrace{\mathcal{E}\left\{ \log_{2}\stackrel[i=1]{M^{N-1}}{\sum}e^{\frac{-\left|\mathbf{h}_{k}\mathbf{W}_{ZF}^{p}\mathbf{x}_{m,i}^{p}\right|^{2}}{2\sigma_{k}^{2}}}\right\} }}.\label{eq:80}
\end{equation}

\noindent By using Jensen inequality, and following similar steps
as in the previous section, we can find the average of $\psi$ as
in (\ref{eq:39-2}).

\subsection{Conventional Transmission NoRS}

The rate at the $k^{th}$ user in conventional transmission without
RS is expressed by

\begin{equation}
\mathcal{E}\left\{ R_{k}^{NoRS}\right\} =\log_{2}M-\,\frac{1}{M^{N}}\stackrel[m=1]{M^{N}}{\sum}\underset{\psi}{\underbrace{\mathcal{E}\left\{ \log_{2}\stackrel[i=1]{M^{N}}{\sum}e^{\frac{-\left|\mathbf{h}_{k}\mathbf{W}\mathbf{x}_{m,i}\right|^{2}}{2\sigma_{k}^{2}}}\right\} }}.\label{eq:35-1}
\end{equation}

 For sake of comparison with using RS technique in this scenario,
we study approximation of the ergodic user rate based on large number
of antennas. In CI case the precoding matrix is given in (\ref{eq:7-2}),
and the expectation in (\ref{eq:35-1}) can be derived using Jensen
inequality as in (\ref{eq:18-2}) and (\ref{eq:72-1}). On the other
hand, in ZF scenario the precoding matrix is given in (\ref{eq:24-1}),
and then the expectation in (\ref{eq:35-1}) can be derived using
Jensen inequality as in (\ref{eq:39-2}).

\section{Rate Maximization through RS Power Allocation}

In this section, we formulate a power allocation problem for maximizing
the ergodic sum-rate of the RS transmission schemes described in the
previous sections. The optimal value of $t$ can be obtained by solving
the following problem

\begin{equation}
\underset{0\leq t\leq1}{\max}\quad\mathcal{E}\left\{ R\right\} \textrm{ in }(\ref{eq:5-1}).\label{eq:73}
\end{equation}

It is worth noting that the availability of perfect CSIT enables the
BS to maximize the instantaneous sum-rate by adapting the power split
among the common and private messages based on the channel status.
Consequently, following \cite{RS5}, the maximization in (\ref{eq:73})
can be moved inside the expectation and the optimum solution can be
found for each channel state. In case the BS has imperfect CSIT, the
BS can not evaluate the instantaneous rates, but it can access the
average rates which are the expected rates for a given channel estimate.
Hence, maximizing the ergodic sum-rate under imperfect CSIT can be
achieved for each estimated channel \cite{RS5}. For simplicity and
to gain some insight, we consider ergodic sum-rate maximization problem
in the two scenarios.

\noindent On one hand, the analytical optimization for the case of
finite constellation signaling using the derived formulas above becomes
intractable. On the other hand, the optimal $t$ can be obtained by
a simple one dimensional search over $0\leq t\leq1$. Hence, the optimal
$t$ can be found by using line search methods such as golden section
technique. The overall steps of golden section method to obtain the
optimal $t$ is stated in Algorithm 1 \cite{algorithm}. 

\begin{algorithm}
\noindent \begin{centering}
Initialize $\varrho=0,\,\zeta=1,\textrm{ and }\lambda=\frac{-1+\sqrt{5}}{2}.$
\par\end{centering}

\noindent \begin{centering}
Repeat
\par\end{centering}

\noindent \begin{centering}
Update $t_{1}=\varrho+(1-\lambda)\zeta\textrm{ and }t_{2}=\zeta+(1-\lambda)\varrho.$
\par\end{centering}

\noindent \begin{centering}
Obtain $R\left(t_{1}\right)$ and $R\left(t_{2}\right)$ from (\ref{eq:5-1}).
\par\end{centering}

\noindent \begin{centering}
If $R\left(t_{1}\right)>R\left(t_{2}\right)$, set $\varrho=t_{1}$.
Else set $\zeta=t_{2}$.
\par\end{centering}

\noindent \begin{centering}
Until $\left|\varrho-\zeta\right|$ converges.
\par\end{centering}

\noindent \begin{centering}
Find $t^{*}=(\varrho+\zeta)/2.$
\par\end{centering}

\protect\caption{Golden Section Method.}
\end{algorithm}

Moreover, in order to reduce the complexity, two sub-optimal solutions
can be considered in finite alphabet scenarios, as follows. 
\begin{itemize}
\item In the first solution, we allocate a fraction $t$ of the total power
for the private messages to achieve the same sum-rate as the conventional
techniques with full power. Then, the remaining power can be allocated
for the common message, as considered in \cite{Rs6}. The sum-rate
payoff of the RS scheme over the NoRS can be determined by,
\end{itemize}
\begin{equation}
\varDelta R=\mathcal{E}\left\{ R_{c}\right\} +\stackrel[k=1]{K}{\sum}\left(\mathcal{E}\left\{ R_{k}^{p}\right\} -\mathcal{E}\left\{ R_{k}^{NoRS}\right\} \right)
\end{equation}

Consequently, the ratio $t$ that achieves the superiority can be
obtained by satisfying the equality, $\mathcal{E}\left\{ R_{k}^{p}\right\} =\mathcal{E}\left\{ R_{k}^{NoRS}\right\} $. 
\begin{itemize}
\item In the second solution, since the achievable data rate in the finite
alphabet systems saturates at maximum predefined value $\left(R_{m}=\left(K+1\right)\log_{2}M\right)$,
here at high SNR the optimal value of $t$ is the value that achieves
the maximum rate with less transmit power $P$, as in the following
expression
\end{itemize}
\begin{equation}
\left(K+1\right)\log_{2}M=\min_{j}\left(\mathcal{E}\left\{ R_{j}^{c}\right\} \right)_{j=1}^{K}+\stackrel[k=1]{K}{\sum}\mathcal{E}\left\{ R_{k}^{p}\right\} \label{eq:76-1}
\end{equation}

Therefore, the optimum value of $t$ at high SNR is the value that
satisfies (\ref{eq:76-1}) with minimum power $P$.

\section{Numerical Results\label{sec:Numerical-Results}}

In this section, we present numerical results of the analytical expressions
derived in this work. Monte-Carlo simulations are conducted where
the channel coefficients are randomly generated. The path loss exponent
is chosen to be $m=2.7$, and assuming the users have same noise power,
$\sigma^{2}$, and the total transmission power is $p$, the transmit
signal to noise ratio (SNR) is defined as SNR = $\frac{p}{\sigma^{2}}$. 

\begin{figure*}
\noindent \begin{centering}
\subfloat[\label{fig:1a}Sum-rate versus SNR, when $d_{1}=d_{2}=1m$.]{\noindent \begin{centering}
\includegraphics[scale=0.55]{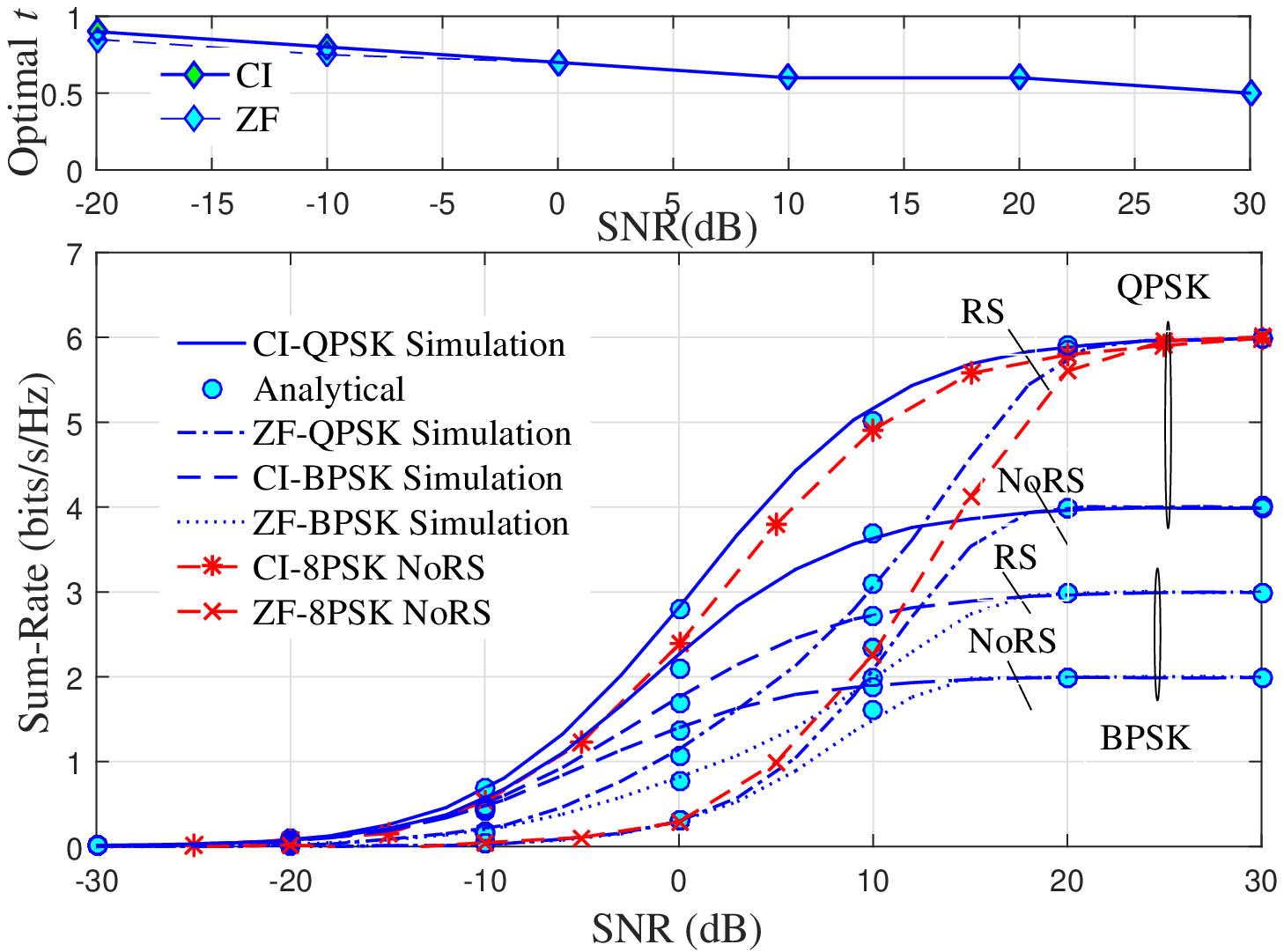}
\par\end{centering}

}\subfloat[\label{fig:1b}Sum-rate versus SNR, when the users are randomly distributed.
]{\noindent \begin{centering}
\includegraphics[scale=0.55]{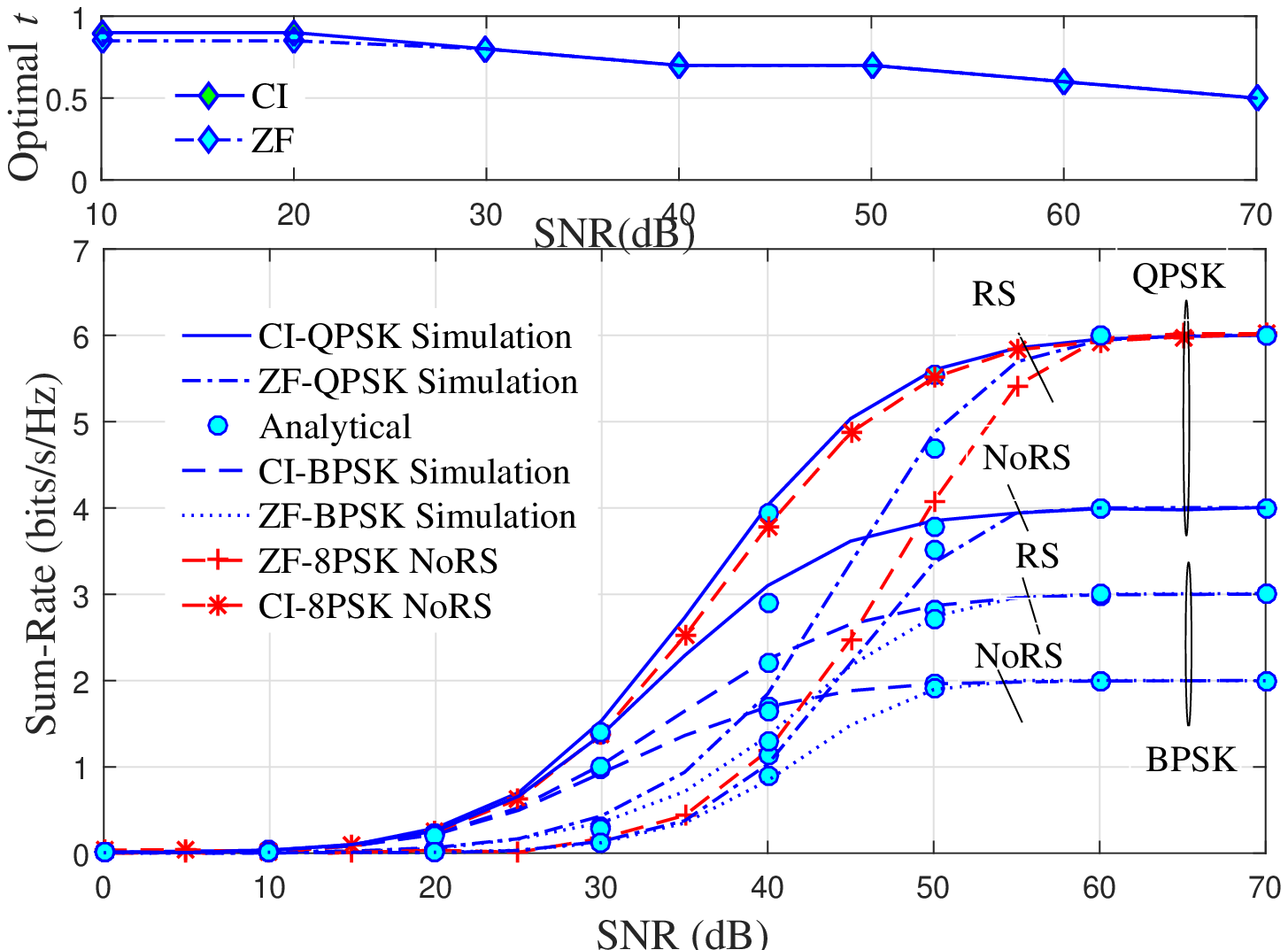}
\par\end{centering}

}
\par\end{centering}

\protect\caption{\label{fig:1}Sum-rate versus SNR for RS and NoRS with different types
of input in perfect CSI, when $N=3\textrm{ and }K=2$. }
\end{figure*}

\begin{figure*}
\begin{centering}
\subfloat[\label{fig:2a}Sum-rate versus SNR, when $d_{1}=d_{2}=1m$.]{\begin{centering}
\includegraphics[scale=0.55]{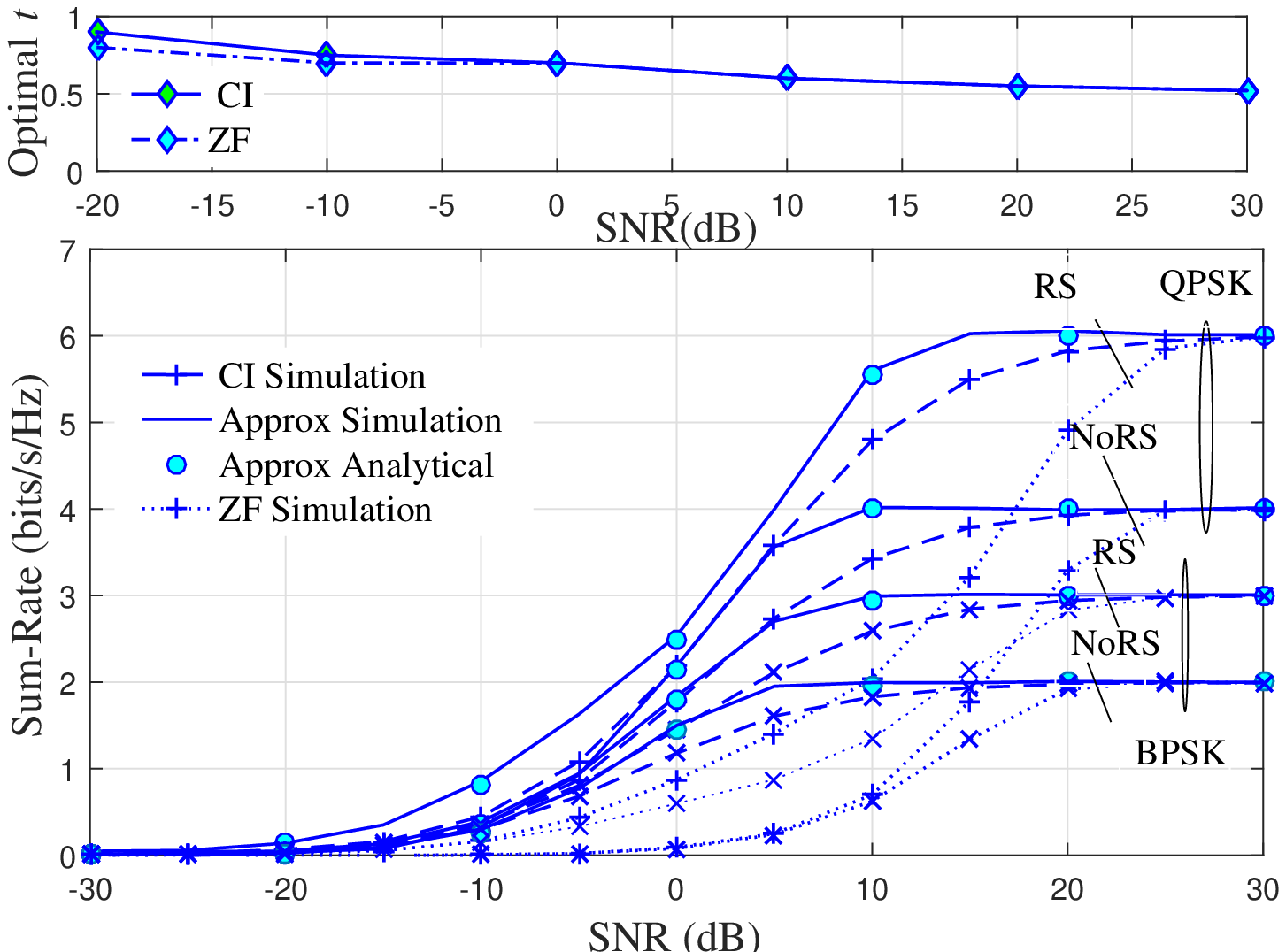}
\par\end{centering}

}\subfloat[\label{fig:2b}Sum-rate versus SNR, when the users are randomly distributed.]{\centering{}\includegraphics[scale=0.55]{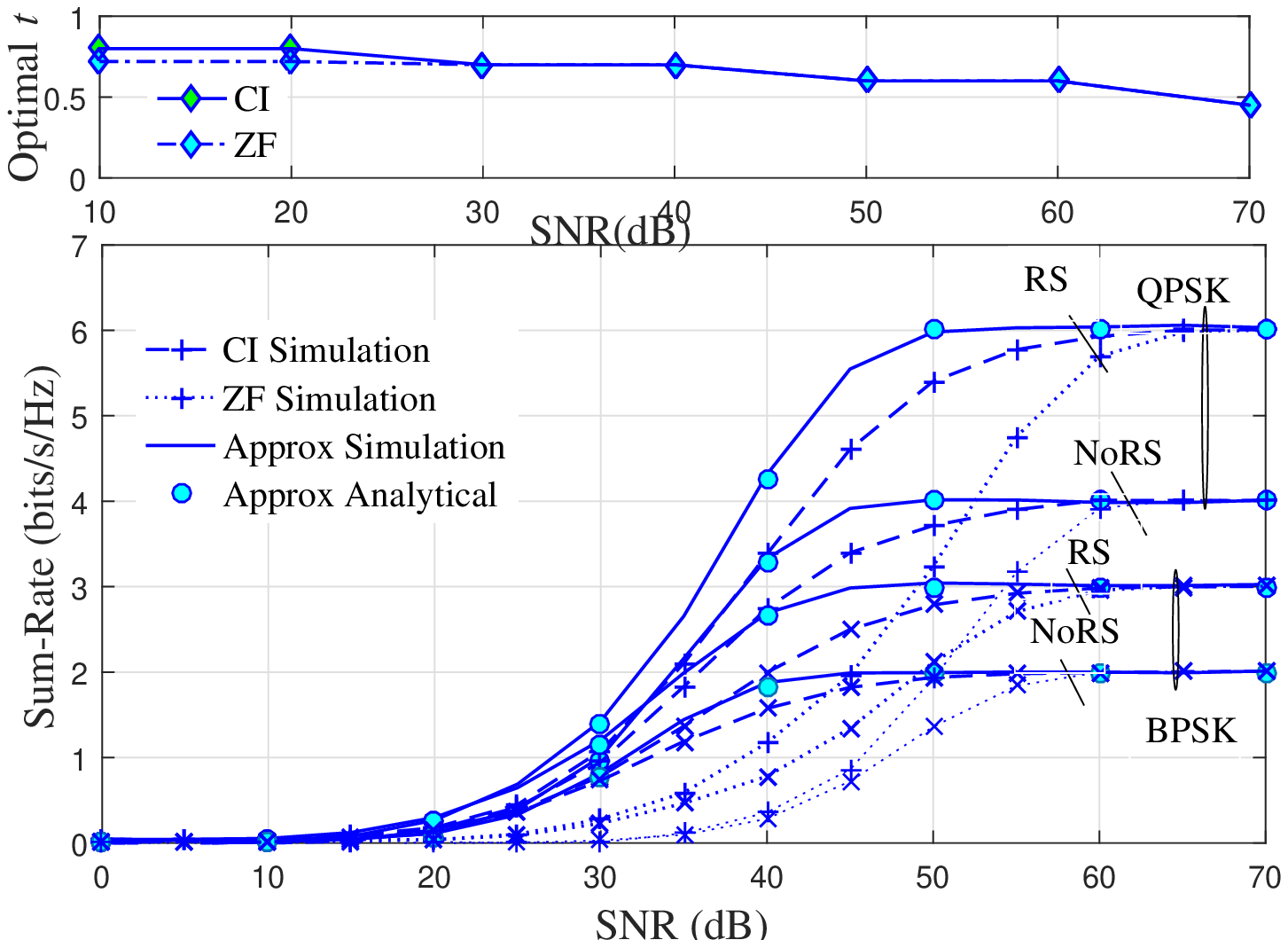}}
\par\end{centering}

\protect\caption{\label{fig:2}Sum-rate versus SNR for RS and NoRS with different types
of input in imperfect CSI, when $N=3\textrm{ and }K=2$.}
\end{figure*}

Firstly, in Fig. \ref{fig:1} and Fig. \ref{fig:2}, we illustrate
the sum-rate for the RS and NoRS using MRT-CI and MRT-ZF in perfect
CSIT scenario and imperfect CSIT scenario, respectively, subject to
BPSK and QPSK when $N=3$, and $K=2$. Fig. \ref{fig:1a} and Fig.
\ref{fig:2a} present the sum-rate in the two scenarios when the distances
between the BS and the users are normalized to unit value, .i.e, without
the impact of the path-loss. Fig. \ref{fig:1b} and Fig. \ref{fig:2b}
show the sum-rate when the users are uniformly distributed inside
a circle area with a radius of 40m and the BS is located at the center
of this area. The good agreement between the analytical and simulated
results confirms the validity of the analysis introduced in this paper.
Several observations can be extracted from these figures. Firstly,
it is clear that the sum rate saturates at a certain SNR value, owing
to the finite constellation. Secondly, the RS scheme enhances the
sum-rate of the considered system and tackles the sum-rate saturation
occurred in the communication systems with PSK signaling. In addition,
it is evident that the CI precoding techniques outperforms the ZF
technique in the all considered scenarios for a wide SNR range with
an up to 10dB gain in the SNR for a given sum rate. Additionally,
in Fig \ref{fig:1} we plot the sum-rate using 8PSK with NoRS, and
observe that the sum-rate in this case saturates at the same rate
as QPSK with RS, .i.e., 6 bits/s/Hz. However, at low SNR, the gain
attained using QPSK with RS is higher than that using 8PSK with NoRS
in all considered schemes. Comparing the results in Fig. \ref{fig:1a}
and Fig. \ref{fig:2a} with that in Fig. \ref{fig:1b} and Fig. \ref{fig:2b},
one can notice that, in general, increasing the distance always degrades
the achievable sum rates. In addition, when the distance between the
BS and the users increases the rate saturation occurs at high SNR
values, due to larger path-loss. It is also clear that, the superiority
of RS with CI over RS with ZF and NoRS does not depend on the users'
locations. Furthermore, as anticipated the system performance degrade
notably in the imperfect CSIT scenario. In addition, we can observe
that when the number of BS antennas is high $N\gg K$, the ZF achieves
the same performance as the CI; ZF precoding can be considered as
a special case of the CI precoding technique \cite{angLi}. 

\begin{figure*}
\noindent \begin{centering}
\subfloat[\label{fig:3a}Sum-rate versus SNR, when $d_{1}=d_{2}=d_{3}=1m$.]{\noindent \begin{centering}
\includegraphics[scale=0.55]{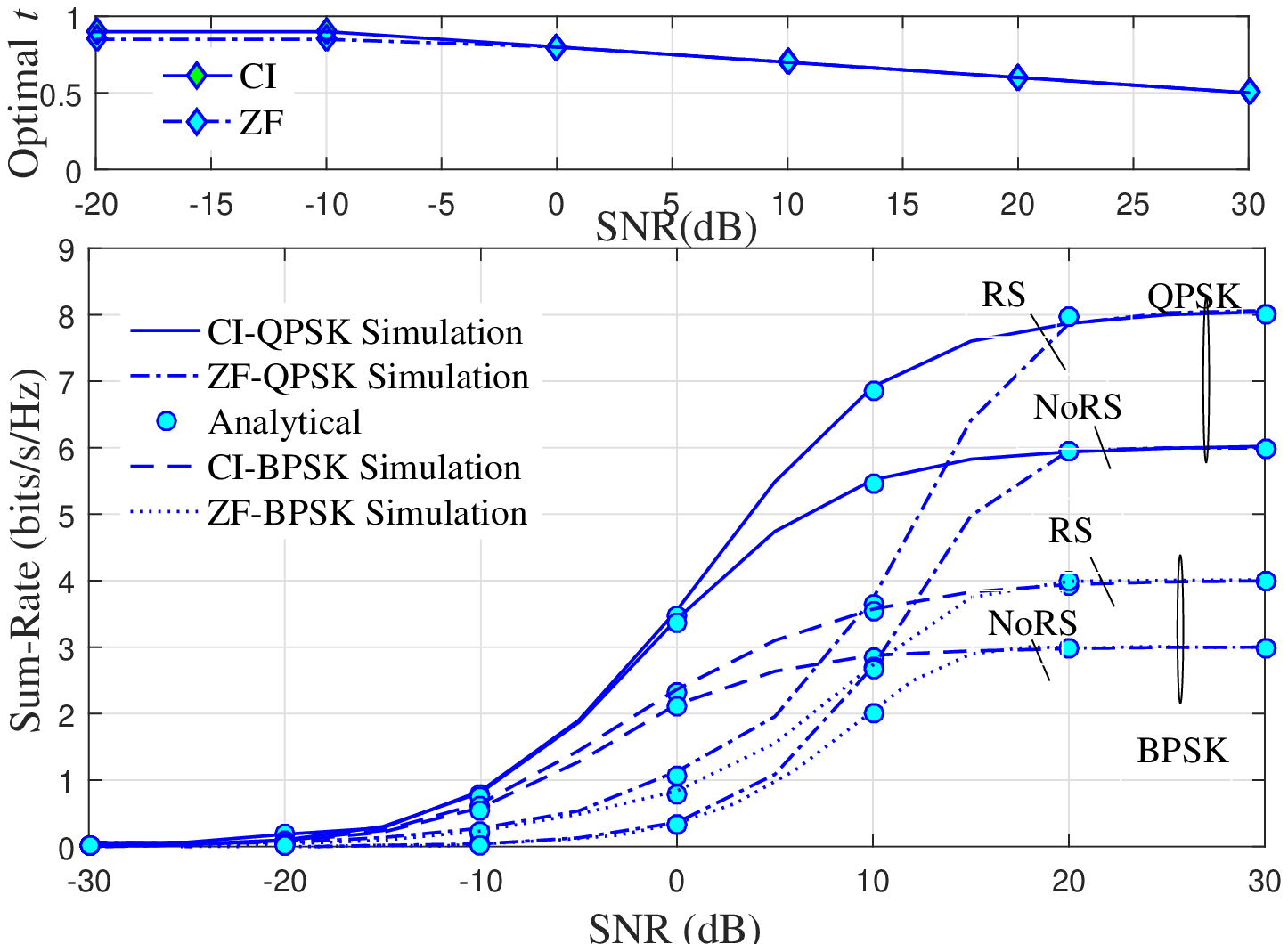}
\par\end{centering}

}\subfloat[\label{fig:3b}Sum-rate versus SNR, when the users are randomly distributed.]{\begin{centering}
\includegraphics[scale=0.55]{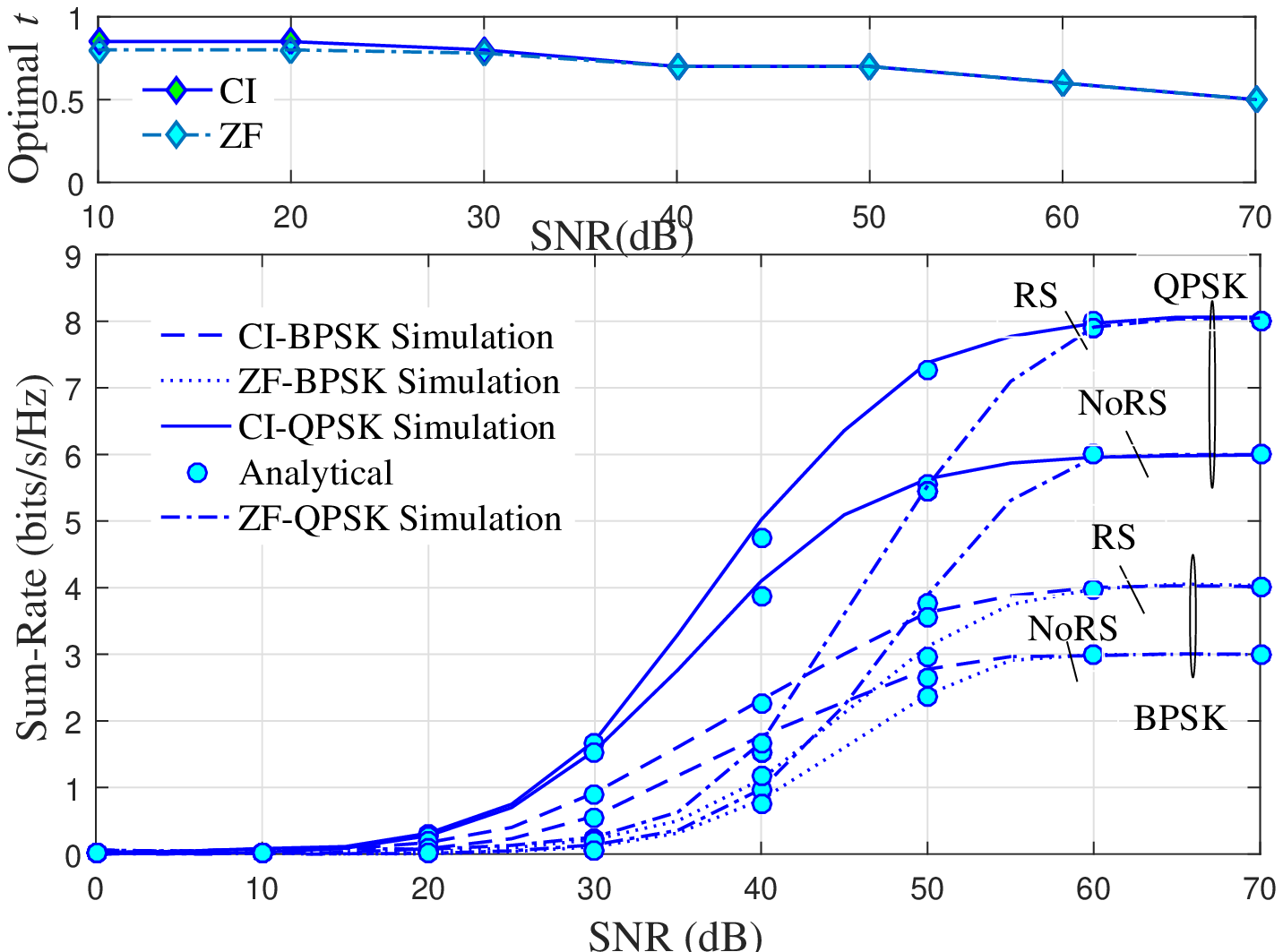}
\par\end{centering}

}
\par\end{centering}

\protect\caption{\label{fig:3}Sum-rate versus SNR for RS and NoRS with different types
of input in perfect CSI, when $N=4\textrm{ and }K=3$.}

\end{figure*}

\begin{figure*}
\noindent \begin{centering}
\subfloat[\label{fig:4a}Sum-rate versus SNR, when $d_{1}=d_{2}=d_{3}=1m$.]{\noindent \begin{centering}
\includegraphics[scale=0.55]{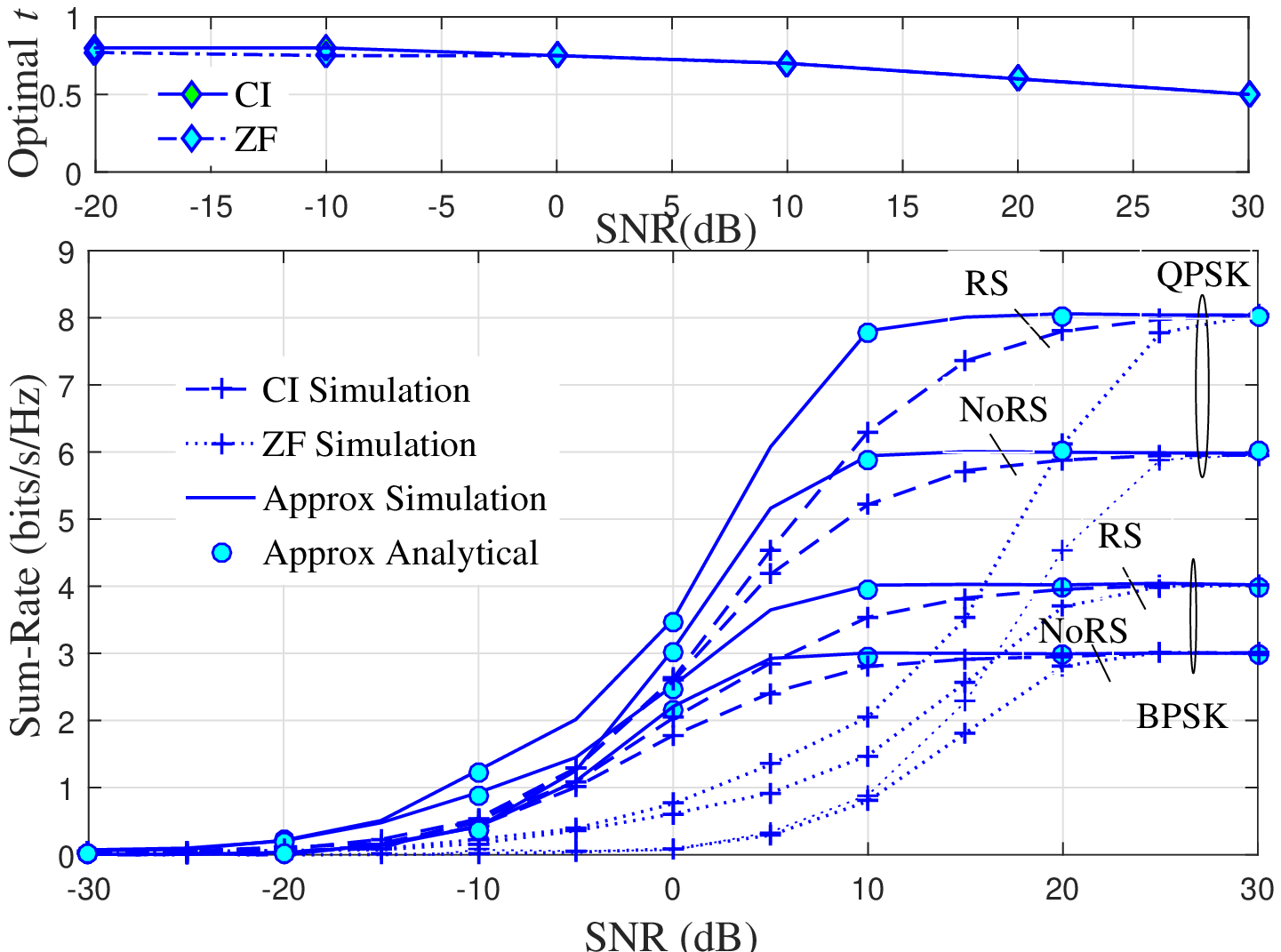}
\par\end{centering}

}\subfloat[\label{fig:4b}Sum-rate versus SNR, when the users are randomly distributed.]{\begin{centering}
\includegraphics[scale=0.55]{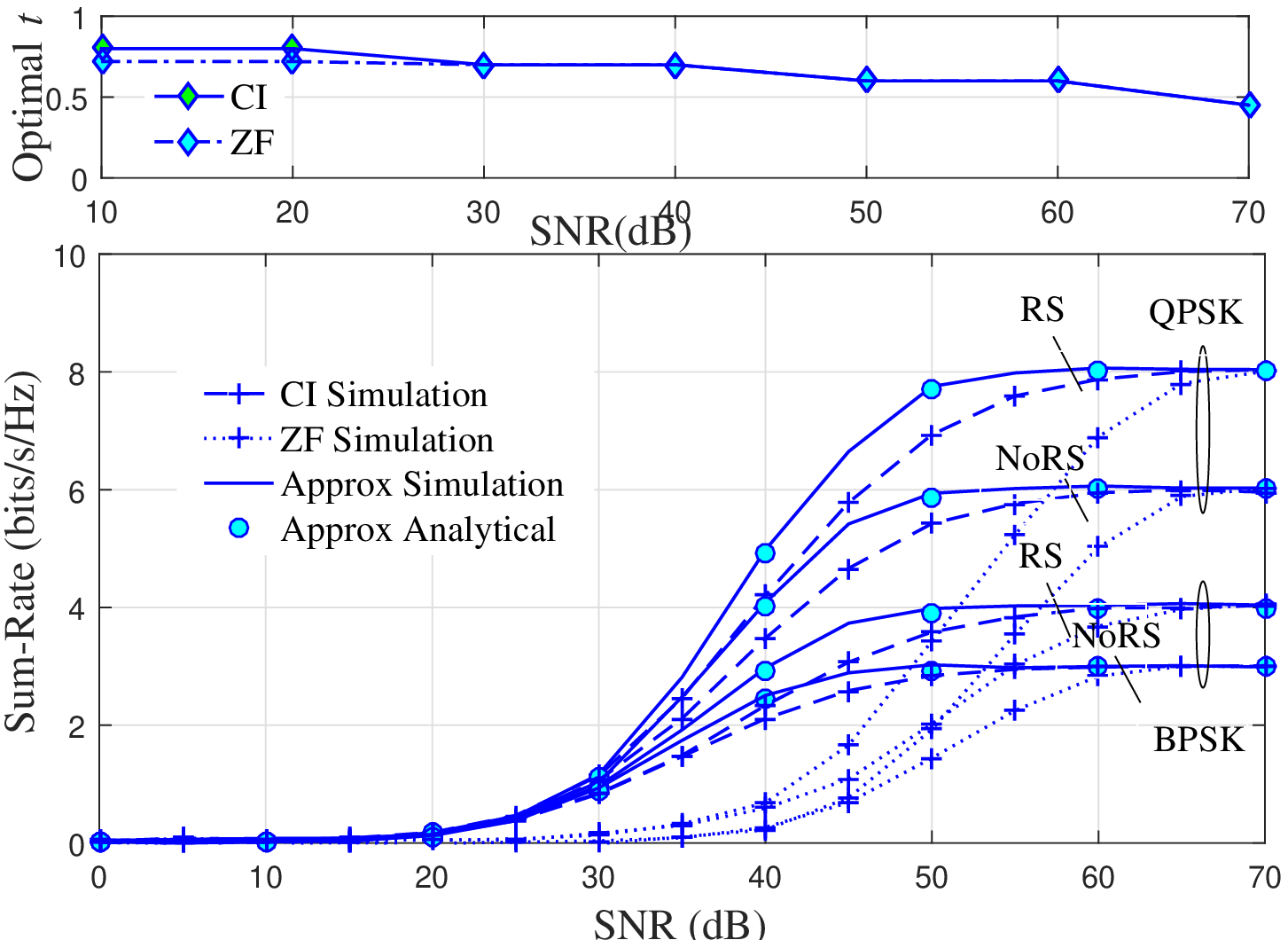}
\par\end{centering}

}
\par\end{centering}

\protect\caption{\label{fig:4}Sum-rate versus SNR for RS and NoRS with different types
of input in imperfect CSI, when $N=4\textrm{ and }K=3$.}
\end{figure*}

Moreover, we investigate the impact of the number of BS antennas and
the number of users on the system performance. Therefore, in Fig.
\ref{fig:3} and Fig. \ref{fig:4} we plot the sum-rate versus the
SNR for the considered transmission schemes with BPSK, and QPSK, when
$N=4$, and $K=3$. Fig. \ref{fig:3a} and Fig. \ref{fig:4a} present
the sum-rate when the distances are normalized to unit value. Fig.
\ref{fig:3b} and Fig. \ref{fig:4b} show the sum-rate when the users
are uniformly distributed in a circle area of 40m radius, where the
BS is located at the center of this area. From the results, it is
clear that increasing the number of users $K$ and/or the number of
antennas $N$ results in enhancing the achievable sum-rate in all
the considered scenarios. In addition, comparing the sum rate achieved
in Fig. \ref{fig:3a} and Fig. \ref{fig:3b}, we can see similar observations
as in the case when $N=3,K=2$. 

Generally, from the results presented in the figures, the optimal
value of the power fraction $t$ at low SNR is approximately $t\approx1$,
which means that splitting the messages and transmitting a common
message is not beneficial in this SNR range. In this case only the
private messages are transmitted and the RS degenerates to NoRS. This
is because the users are experienced similar SNR. If there is a notable
disparity of channel strengths among users, this conclusion may not
hold \cite{BrunoB}. On the other hand at high SNR the optimal value
of $t$ is less than one, $t<1$, which indicates that the common
message is transmitted with the remaining power beyond the saturation
of the private message transmission.

\begin{figure}
\noindent \begin{centering}
\includegraphics[scale=0.55]{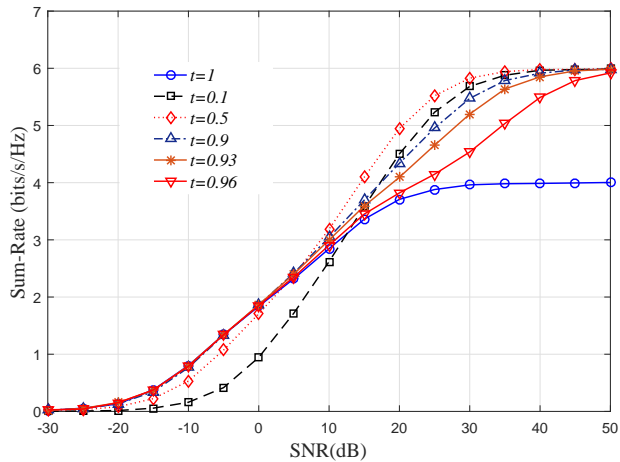}
\par\end{centering}

\protect\caption{\label{fig:5}Sum-rate for RS with CI and QPSK modulation versus SNR
for various values of $t$.}

\end{figure}

In order to clearly illustrate the impact of the power fraction $t$
on the system performance, we plot in Fig. \ref{fig:5} the sum-rate
versus SNR for various values of $t$ with the CI precoding under
QPSK, when $N=3,\, K=2,$ $d_{1}=1$m and $d_{2}=5$m.\textcolor{blue}{{}
}Interestingly enough, it is noted that at low SNR values, $\textrm{SNR}\leq12\textrm{ dB}$,
the sum-rate degrades as $t$ becomes small, and the optimal $t$
in this range is approximately close to 1. In addition, at high SNR
values, $\textrm{SNR}\geq12\textrm{ dB}$, the sum-rate degrades as
the value of $t$ increases, till the sum-rate reaches the achievable
rate in case NoRS when $t=1$.

\section{Conclusions\label{sec:Conclusions}}

In this paper we employed the CI precoding technique to enhance the
sum-rate performed by RS scheme in MU-MIMO systems under PSK input
alphabet. New analytical expressions for the ergodic sum-rate have
been derived for CI precoding technique and ZF precoding technique
in RS and NoRS scenarios. Furthermore, a power allocation scheme that
achieves superiority of RS over NoRS in the presence of finite constellation
was proposed. The results presented in this work demonstrated that
RS with CI has greater sum-rate than RS with ZF and NoRS transmission
techniques. In addition, increasing the number of BS antennas and/
or the number of users enhances the achievable sum-rate. 

\bibliographystyle{IEEEtran}
\bibliography{bib}

\begin{thebibliography}{10}
\providecommand{\url}[1]{#1}
\csname url@samestyle\endcsname
\providecommand{\newblock}{\relax}
\providecommand{\bibinfo}[2]{#2}
\providecommand{\BIBentrySTDinterwordspacing}{\spaceskip=0pt\relax}
\providecommand{\BIBentryALTinterwordstretchfactor}{4}
\providecommand{\BIBentryALTinterwordspacing}{\spaceskip=\fontdimen2\font plus
\BIBentryALTinterwordstretchfactor\fontdimen3\font minus
  \fontdimen4\font\relax}
\providecommand{\BIBforeignlanguage}[2]{{%
\expandafter\ifx\csname l@#1\endcsname\relax
\typeout{** WARNING: IEEEtran.bst: No hyphenation pattern has been}%
\typeout{** loaded for the language `#1'. Using the pattern for}%
\typeout{** the default language instead.}%
\else
\language=\csname l@#1\endcsname
\fi
#2}}
\providecommand{\BIBdecl}{\relax}
\BIBdecl

\bibitem{WCNC19}
A.~Salem and C.~Masouros, ``Rate splitting approach under psk signaling using
  constructive interference precoding technique,'' in \emph{Proc. IEEE Wireless
  Commun. Netw. Conf. (WCNC)}, 2019.

\bibitem{MIMO1}
M.~S. John G.~Proakis, \emph{Digital Communications, Fifth Edition}.\hskip 1em
  plus 0.5em minus 0.4em\relax McGraw-Hill, NY USA, 2008.

\bibitem{MIMO2}
C.~B.~P. Howard~Huang and S.~Venkatesan, \emph{MIMO Communication for cellular
  Networks}.\hskip 1em plus 0.5em minus 0.4em\relax Springer, 2012, 2008.

\bibitem{MIMO3}
Y.~Wu, C.~Xiao, X.~Gao, J.~D. Matyjas, and Z.~Ding, ``Linear precoder design
  for mimo interference channels with finite-alphabet signaling,'' \emph{IEEE
  Transactions on Communications}, vol.~61, no.~9, pp. 3766--3780, September
  2013.

\bibitem{pdfZF}
D.~Lee, ``Performance analysis of zero-forcing-precoded scheduling system with
  adaptive modulation for multiuser-multiple input multiple output
  transmission,'' \emph{IET Communications}, vol.~9, no.~16, pp. 2007--2012,
  2015.

\bibitem{Twway}
A.~Salem and K.~A. Hamdi, ``Wireless power transfer in multi-pair two-way af
  relaying networks,'' \emph{IEEE Transactions on Communications}, vol.~64,
  no.~11, pp. 4578--4591, Nov 2016.

\bibitem{RS1}
B.~Clerckx, H.~Joudeh, C.~Hao, M.~Dai, and B.~Rassouli, ``Rate splitting for
  mimo wireless networks: a promising phy-layer strategy for lte evolution,''
  \emph{IEEE Communications Magazine}, vol.~54, no.~5, pp. 98--105, May 2016.

\bibitem{Rs3}
H.~Joudeh and B.~Clerckx, ``Robust transmission in downlink multiuser miso
  systems: A rate-splitting approach,'' \emph{IEEE Transactions on Signal
  Processing}, vol.~64, no.~23, pp. 6227--6242, Dec 2016.

\bibitem{Rs2}
C.~Hao, Y.~Wu, and B.~Clerckx, ``Rate analysis of two-receiver miso broadcast
  channel with finite rate feedback: A rate-splitting approach,'' \emph{IEEE
  Transactions on Communications}, vol.~63, no.~9, pp. 3232--3246, Sept 2015.

\bibitem{Rs4}
A.~Papazafeiropoulos and T.~Ratnarajah, ``Rate-splitting robustness in
  multi-pair massive mimo relay systems,'' \emph{IEEE Transactions on Wireless
  Communications}, vol.~17, no.~8, pp. 5623--5636, Aug 2018.

\bibitem{RS5}
H.~Joudeh and B.~Clerckx, ``Sum-rate maximization for linearly precoded
  downlink multiuser miso systems with partial csit: A rate-splitting
  approach,'' \emph{IEEE Transactions on Communications}, vol.~64, no.~11, pp.
  4847--4861, Nov 2016.

\bibitem{Rs6}
M.~Dai, B.~Clerckx, D.~Gesbert, and G.~Caire, ``A rate splitting strategy for
  massive mimo with imperfect csit,'' \emph{IEEE Transactions on Wireless
  Communications}, vol.~15, no.~7, pp. 4611--4624, July 2016.

\bibitem{newrs2}
C.~Hao and B.~Clerckx, ``Miso networks with imperfect csit: A topological
  rate-splitting approach,'' \emph{IEEE Transactions on Communications},
  vol.~65, no.~5, pp. 2164--2179, May 2017.

\bibitem{BrunoA}
H.~{Joudeh} and B.~{Clerckx}, ``Rate-splitting for max-min fair multigroup
  multicast beamforming in overloaded systems,'' \emph{IEEE Transactions on
  Wireless Communications}, vol.~16, no.~11, pp. 7276--7289, Nov 2017.

\bibitem{BrunoB}
\BIBentryALTinterwordspacing
Y.~Mao, B.~Clerckx, and V.~O. Li, ``Rate-splitting multiple access for downlink
  communication systems: bridging, generalizing, and outperforming sdma and
  noma,'' \emph{EURASIP Journal on Wireless Communications and Networking},
  vol. 2018, no.~1, p. 133, May 2018. [Online]. Available:
  \url{https://doi.org/10.1186/s13638-018-1104-7}
\BIBentrySTDinterwordspacing

\bibitem{CI1}
C.~Masouros and E.~Alsusa, ``Dynamic linear precoding for the exploitation of
  known interference in mimo broadcast systems,'' \emph{IEEE Transactions on
  Wireless Communications}, vol.~8, no.~3, pp. 1396--1404, March 2009.

\bibitem{A}
C.~Masouros, M.~Sellathurai, and T.~Ratnarajah, ``Vector perturbation based on
  symbol scaling for limited feedback miso downlinks,'' \emph{IEEE Transactions
  on Signal Processing}, vol.~62, no.~3, pp. 562--571, Feb 2014.

\bibitem{CI2}
C.~Masouros and G.~Zheng, ``Exploiting known interference as green signal power
  for downlink beamforming optimization,'' \emph{IEEE Transactions on Signal
  Processing}, vol.~63, no.~14, pp. 3628--3640, July 2015.

\bibitem{CI3}
S.~Timotheou, G.~Zheng, C.~Masouros, and I.~Krikidis, ``Exploiting constructive
  interference for simultaneous wireless information and power transfer in
  multiuser downlink systems,'' \emph{IEEE Journal on Selected Areas in
  Communications}, vol.~34, no.~5, pp. 1772--1784, May 2016.

\bibitem{Luxm0}
M.~Alodeh, S.~Chatzinotas, and B.~Ottersten, ``Constructive multiuser
  interference in symbol level precoding for the miso downlink channel,''
  \emph{IEEE Transactions on Signal Processing}, vol.~63, no.~9, pp.
  2239--2252, May 2015.

\bibitem{Luxm1}
A.~Haqiqatnejad, F.~Kayhan, and B.~Ottersten, ``Symbol-level precoding design
  based on distance preserving constructive interference regions,'' \emph{IEEE
  Transactions on Signal Processing}, vol.~66, no.~22, pp. 5817--5832, Nov
  2018.

\bibitem{Luxm2}
------, ``Constructive interference for generic constellations,'' \emph{IEEE
  Signal Processing Letters}, vol.~25, no.~4, pp. 586--590, April 2018.

\bibitem{CI5}
M.~R.~A. Khandaker, C.~Masouros, and K.~K. Wong, ``Constructive interference
  based secure precoding: A new dimension in physical layer security,''
  \emph{IEEE Transactions on Information Forensics and Security}, vol.~13,
  no.~9, pp. 2256--2268, Sept 2018.

\bibitem{CI4}
P.~V. Amadori and C.~Masouros, ``Large scale antenna selection and precoding
  for interference exploitation,'' \emph{IEEE Transactions on Communications},
  vol.~65, no.~10, pp. 4529--4542, Oct 2017.

\bibitem{angLi}
A.~Li and C.~Masouros, ``Interference exploitation precoding made practical:
  Optimal closed-form solutions for psk modulations,'' \emph{IEEE Transactions
  on Wireless Communications}, pp. 1--1, 2018.

\bibitem{bookaspects}
R.~J. Muirhead, \emph{Aspects of Multivariate Statistical Theory}, 1982.

\bibitem{interference1}
W.~Wu, K.~Wang, W.~Zeng, Z.~Ding, and C.~Xiao, ``Cooperative multi-cell mimo
  downlink precoding with finite-alphabet inputs,'' \emph{IEEE Transactions on
  Communications}, vol.~63, no.~3, pp. 766--779, March 2015.

\bibitem{precoding6}
Y.~Wu, C.~Xiao, X.~Gao, J.~D. Matyjas, and Z.~Ding, ``Linear precoder design
  for mimo interference channels with finite-alphabet signaling,'' \emph{IEEE
  Transactions on Communications}, vol.~61, no.~9, pp. 3766--3780, September
  2013.

\bibitem{precoding2}
Y.~Wu, M.~Wang, C.~Xiao, Z.~Ding, and X.~Gao, ``Linear precoding for mimo
  broadcast channels with finite-alphabet constraints,'' \emph{IEEE
  Transactions on Wireless Communications}, vol.~11, no.~8, pp. 2906--2920,
  August 2012.

\bibitem{book2}
M.~Abramowitz and I.~A. Stegun, \emph{Handbook of Mathematical Functions With
  Formulas, Graphs, and Mathematical Tabl}, Washington,D.C.: U.S. Dept.
  Commerce, 1972.

\bibitem{book}
------, \emph{Handbook of Mathematical Functions With Formulas, Graphs, and
  Mathematical Tabl}, Washington,D.C.: U.S. Dept. Commerce, 1972.

\bibitem{marzita}
H.~Q. Ngo, E.~G. Larsson, and T.~L. Marzetta, ``Energy and spectral efficiency
  of very large multiuser mimo systems,'' \emph{IEEE Transactions on
  Communications}, vol.~61, no.~4, pp. 1436--1449, April 2013.

\bibitem{salim}
M.~. Alouini and A.~J. Goldsmith, ``Area spectral efficiency of cellular mobile
  radio systems,'' \emph{IEEE Transactions on Vehicular Technology}, vol.~48,
  no.~4, pp. 1047--1066, July 1999.

\bibitem{algorithm}
J.~{Kim}, H.~{Lee}, C.~{Song}, T.~{Oh}, and I.~{Lee}, ``Sum throughput
  maximization for multi-user mimo cognitive wireless powered communication
  networks,'' \emph{IEEE Transactions on Wireless Communications}, vol.~16,
  no.~2, pp. 913--923, Feb 2017.

\end{thebibliography}

\end{document}